\documentclass[12pt,fleqn]{article}

\usepackage{amsmath, amssymb,graphicx}

\renewcommand{\a}{\alpha}
\renewcommand{\b}{\beta}
  \newcommand{\g}{\gamma}
\renewcommand{\d}{\delta}
  \newcommand{\s}{\sigma}

\renewcommand{\l}{\lambda}
  \newcommand{\m}{\mu}
  \newcommand{\n}{\nu}
  \newcommand{\ve}{\varepsilon}

\def\di{{\rm d}}
\def\ts{\textstyle}
  \newcommand{\N}{\mathcal{N}}

\newcommand{\Qsl}{Q\!\!\!\!\slash\,}
\newcommand{\Dsl}{D\!\!\!\!\slash\,}
\newcommand{\thetasl}{\theta\!\!\!\slash}

\def\Bpm{B^{\! \pm}}
\def\Bp{B^{\! +}}
\def\Bm{B^{\! -}}
\def\al{\alpha}
\def\pr{\partial}

\begin{document}
\bigskip

\title{
{\bf Four-Dimensional Superconformal Theories with Interacting 
Boundaries or  Defects}\vspace{2cm}}
\bigskip

\medskip

\author{{\bf Johanna Erdmenger, Zachary Guralnik and Ingo Kirsch} 
\thanks{email:
    jke@physik.hu-berlin.de, zack@physik.hu-berlin.de, 
    ik@physik.hu-berlin.de}\\ \\ \\
  \\ Institut f\"ur Physik\\ 
  Humboldt-Universit\"at zu Berlin\\ 
Invalidenstra{\ss}e 110\\
D-10115 Berlin, Germany }

\date{}

\maketitle

\begin{abstract} { 
 
 We study four-dimensional superconformal field theories coupled to 
 three-dimen\-sional superconformal boundary or defect degrees of
 freedom. Starting with bulk ${\N}=2$, $d=4$ theories, we construct
 abelian models preserving ${\N}=2$, $d=3$ supersymmetry and the
 conformal symmetries under which the boundary/defect is invariant.
 We write the action, including the bulk terms, in ${\N}=2$, $d=3$ 
 superspace. Moreover we derive Callan-Symanzik equations for these 
 models using their superconformal transformation properties and show 
 that the beta functions vanish to all orders in perturbation theory,
 such that the models remain superconformal upon quantization. 
 Furthermore we study a model with $\N=4$ $SU(N)$ Yang-Mills theory in the
 bulk coupled to a $\N=4$, $d=3$ hypermultiplet on a defect.  This model was
 constructed by DeWolfe, Freedman and Ooguri, and conjectured to be conformal
 based on its relation to an AdS configuration stu\-died by Karch and Randall.
 We write this model in $\N=2$, $d=3$ superspace, which has the
 distinct advantage that non-renormalization theorems become transparent.
 Using $\N=4$, $d=3$ supersymmetry, we argue that the model is conformal.

}

\smallskip

\hfill HU-EP-02/07 

\end{abstract}

\thispagestyle{empty}

\newpage
\section{Introduction}
\setcounter{equation}{0}

Conformal field theories in $d$ dimensions with a boundary of codimension one
have interesting general properties, which have been
investigated in \cite{Cardy,OsbornMcAvity}. The
essential feature of such field theories is that the conformal group in the
$d$ dimensional space is broken from $SO(d,2)$ down to $SO(d-1,2)$ in the
presence of the boundary.  The unbroken conformal generators are those which
leave the boundary invariant.  In \cite{OsbornMcAvity}, correlation functions
for general boundary conformal field theories where constructed by symmetry
considerations and by deriving a boundary operator product expansion. Due to
the reduced conformal symmetry, these correlation functions are generally less
constrained than in conformal field theories without a boundary.

There are examples for 
conformal field theories of this type in which there are additional
degrees of freedom at a defect or boundary.
Such impurity theories were studied for instance in the context of matrix
descriptions of compactified five-branes
\cite{Sethi,GanorSethi,KapustinSethi}, and also in the context of the
AdS/CFT correspondence.
Recently, Karch and Randall \cite{Karch} proposed an AdS/CFT duality 
for D3/D5 brane systems whose near-horizon limit is $AdS_5 \times S^5$ 
with D5 branes wrapping
an $AdS_4 \times S^2$ submanifold.  They conjecture the dual field theory to
be a four-dimensional ${\cal N}=4$ Yang-Mills theory interacting with a
three-dimensional conformal field theory in such a way as to preserve the
common conformal symmetries.  The three-dimensional degrees of freedom were
proposed to be the holographic description of modes living on the D5-brane on
which there is ``locally localized gravity''.

Within this context, there are two different scenarios: In the first,
all of the D3 branes intersect the D5 branes. The dual field theory is then
expected to contain a defect on which the three-dimensional theory lives.  In
the second scenario, some of the D3 branes end on the D5 branes, allowing for
the interesting possibility of having two four-dimensional conformal field
theories with different central charges coupled at a common boundary to a
three-dimensional conformal field theory.

In the defect scenario, the AdS/CFT correspondence was subsequently
investigated in detail by DeWolfe, Freedman  and Ooguri \cite{Freedman}.
These authors explicitly construct the Lagrangian of the dual field theory,
which at the classical level preserves a $SO(3,2)$ conformal symmetry.  The
bulk component is a $\N=4$, $d=4$ super Yang-Mills theory, half the
modes of which are coupled to a defect $\N =4$, $d=3$ hypermultiplet.  In the
construction of \cite{Freedman}, the bulk modes are coupled to the defect modes
in a manner preserving half the bulk supersymmetries by defining the defect in
$\N=1$, $d=4$  superspace. In analogy to earlier results in
\cite{Hori}, the defect locus is written as a condition on both
a spatial coordinate, $x_2=0$, and the Grassmann coordinate, $\theta = {\bar
  \theta}$.  Evaluated at the defect, bulk $\N=1$ superfields become $\N=1$,
$d=3$ superfields which can be directly coupled to defect degrees of freedom.
The $SO(3,2)$ symmetries of the supergravity dual strongly suggest that the
conformal invariance of the classical theory is preserved by quantum
corrections.   There are also arguments in the context of matrix theory which
suggest conformal invariance \cite{GanorSethi,sethicom}. 
Partial field-theoretical arguments for conformal invariance  were given in
\cite{Freedman}, and a proof was given for the abelian version of the model
which has no bulk interactions.

In this paper we shall revisit the defect model considered in
\cite{Freedman}.
We also construct
other similar models preserving at least $\N=2$, $d=3$
supersymmetry. In addition to the defect case, where there are no boundary
conditions, we also consider boundaries with suitable supersymmetric boundary
conditions.  We shall write both bulk and defect/boundary terms in $\N=2$,
$d=3$ superspace.  In addition to being compact and making many of the
unbroken symmetries manifest, this notation has the distinct advantage that
non-renormalization theorems are more transparent due to the existence of
chiral superfields not present in ${\cal N}=1$, $d=3$ language.  Furthermore
in $\N=2$, $d=3$ language it is easy to write Feynman graphs with
bulk-boundary interactions.  A similiar procedure for coupling four
dimensional supersymmetric actions to higher dimensional ones was developed in
\cite{HG,Hebecker} in the context of phenomenological model building.

We begin by considering an abelian bulk $\N=2$ vector multiplet with half the
degrees of freedom coupled to charged $\N=2$, $d=3$ chiral multiplets at a
defect (or boundary) in such a way that $\N=2$, $d=3$ superconformal
invariance is classically preserved. For the boundary case, we obtain an
additional $d=3$ Chern-Simons term as a boundary term of the $d=4$ action.  We
then derive a Callan-Symanzik equation by considering the superconformal
transformation properties of the 1PI action in $\N=2$, $d=3$ superspace.  In
the abelian case, the bulk contribution to the action is free, and when
studying the renormalization properties of the 1PI action, it is sufficient to
consider the $d=3$ theory since all vertices live on the boundary or defect.
The Callan-Symanzik equation enables us to show that the beta function
vanishes to all orders in perturbation theory, such that the $\N =2$, $d=3$
superconformal symmetry is preserved by quantum corrections.  A crucial
ingredient in the proof of quantum conformal invariance is the absence of
quantum corrections involving the Chern-Simons term. Such a term cannot
contribute to the local superconformal transformation of the quantum action
since its local form is not gauge invariant. This implies the absence
of the gauge beta function. 
Nevertheless the boundary or defect fields acquire an anomalous
dimension in this $\N=2$ model, which does not affect superconformal
invariance. 

In the defect case no boundary conditions are imposed on bulk fields, whereas
in the boundary case we impose Neumann boundary conditions, which - in
contrast to Dirichlet conditions - allow for coupling the electrically
charged boundary degrees of freedom to the bulk fields.
We expect mirror symmetric models with Dirichlet boundary conditions
to exist as well.
As far as the conformal invariance of
the models considered here is concerned, it does not matter whether one has a
defect or a boundary. 
We emphasize that, unlike the defect model, the boundary model we construct
does not correspond to a D3/D5 system, which would require Dirichlet boundary
conditions \cite{HananyWitten}.  It is nevertheless interesting as a toy model
and is a first step towards considering models in which there are
different bulk central charges on opposite sides of the
boundary.  Such conformal field theories might be expected to exist as
holographic duals of supergravity configurations discussed in \cite{Karch} in
which two $AdS_5$ backgrounds with different curvature are separated by an
$AdS_4$ submanifold.

We also consider the abelian defect model of \cite{Freedman}, as well
as its boundary version. This model is a bulk
$\N=4$ theory with half the degrees of freedom coupled to a charged $\N=4$,
$d=3$ hypermultiplet living on the defect.  Using $\N=2$, $d=3$ superspace, we
derive the Callan-Symanzik equation for this model and show that the beta
functions and anomalous dimensions vanish. In \cite{Freedman}, a similar proof
was given in component language using power counting and symmetry 
arguments\footnote{In \cite{Freedman}, it 
was argued that the quantum corrections to
  the defect field propagators give rise to divergences which are at most
  logarithmic, such that the defect fields acquire anomalous dimensions. Using
  our $\N=2$ superspace approach, we are in fact able to show that for
 the elementary defect fields, even the
  logarithmic divergences are absent, such that these fields do not
  acquire anomalous dimensions. However composite operators
  may still have anomalous dimensions, which we do not
  consider here.}.

Finally we consider the non-abelian version of the defect model of
\cite{Freedman}, whose conformal invariance has not been previously 
demonstrated.  In the non-abelian case, the analysis of potential
quantum corrections is more involved, since the bulk
action is no longer free. Assuming unbroken $\N =4$, $d=3$ supersymmetry, we
argue that the beta functions of this theory vanish as well.

The paper is organized as follows. In section 2 we discuss the embedding of
\mbox{$\N=2$}, \mbox{$d=3$} superspace in $\N=2$, $d=4$ superspace in the
presence of a boundary or defect.  Moreover we decompose the $\N=2$, $d=4$
vector multiplet under $\N=2$, $d=3$ supersymmetry.  In section 3 we construct
the action for a free abelian $\N=2$, $d=4$ vector multiplet in the bulk
coupled to a charged $\N=2$, $d=3$ chiral multiplet on a boundary or defect.
We investigate the superconformal transformation properties of the quantized
version of this model, derive its Callan-Symanzik equation and show that its
beta function vanishes.  In section 4, we consider the model of
\cite{Freedman} with $\N=4$ super Yang-Mills theory in the bulk coupled to a
$\N=4$, $d=3$ charged hypermultiplet at the defect or boundary. For the
abelian version of this model we show that this model is not renormalized,
using the superconformal transformation properties of the 1PI action again.
In section 5 we consider the non-abelian version of this model and demonstrate
its conformal invariance assuming unbroken $\N=4$, $d=3$ supersymmetry.  We
conclude in section 6.

\section{Decomposing $\N=2$, $d=4$ multiplets under $\N=2$, $d=3$ 
supersymmetry} \setcounter{equation}{0}

Our aim is to couple four-dimensional theories with $\N=2$ or $\N=4$
supersymmetry to a three-dimensional boundary theory at $x_2=0$.  
The super Poincar\'{e} symmetries of the four-dimensional bulk  are 
broken by boundary conditions and defect or boundary  couplings.
For the purpose of coupling the bulk and boundary or defect actions, and for 
computing quantum corrections,  it is convenient to write the 
four-dimensional bulk contribution to the action in a language in which
only the preserved $\N=2$, $d=3$  symmetry is
manifest\footnote{An analogous procedure was considered in 
  \cite{HG} in coupling four-dimensional 
boundary theories to five dimensional bulk theories.}.   To this end
it is necessary to know the decomposition of the higher  dimensional
multiplets under the lower dimensional  supersymmetry.

\subsection{Embedding $\N=2$, $d=3$ in $\N=2$, $d=4$}

We begin by showing how to embed $\N=2$, $d=3$ superspace into $\N=2$, $d=4$
superspace.  For this purpose we perform a twofold coordinate transformation
in $\N=2$, $d=4$ superspace,
\begin{align}
(x,\theta_1,\bar \theta^1, \theta_2,\bar \theta^2) \rightarrow (x,\tilde
\theta_1,\tilde \theta_2,\tilde \thetasl_1,\tilde \thetasl_2) \rightarrow
(x,\theta,\bar \theta, \thetasl,\bar \thetasl)\,. \label{thetatr}
\end{align}
First, we define real spinors, $\tilde \theta_{i}$ and $\tilde
\thetasl_{i}$, as
\begin{align} \label{theta12}
  \tilde \theta_{i} \equiv \textstyle\frac{1}{2} (\theta_{i} + \bar \theta^{i}
  )\,,\quad \tilde \thetasl{}_{i} \equiv \textstyle\frac{1}{2i} (\theta_{i} -
  \bar \theta^{i})\,, \quad i=1,2\,.
\end{align}
Each real spinor is an irreducible representation of the three-dimensional
Lorentz group $SU(1,1) \simeq \overline{SL}(2,\mathbb{R}) \simeq
\overline{SO}(1,2)$.
Subsequently, we rearrange them in the complex spinors $\theta$ and
$\thetasl$,   
\begin{align} \label{trafo2}
  \theta \equiv \textstyle \tilde \theta_{1} - i \tilde
  \theta_{2}, \quad \thetasl \equiv \textstyle \tilde
  \thetasl{}_{1} - i \tilde \thetasl{}_{2} \,.
\end{align}
As we will see shortly,  setting $\thetasl =0$ yields a $\N=2$, $d=3$ 
superspace. 

In the absence of central charges, the $\N=4$, $d=2$ supersymmetry algebra is
\begin{align}
\{Q_{i\a}, \bar Q^{j}{}_{\dot\b}\} &= 2 \s^\m_{\a\dot\b} P_\m \d_{i}^j, 
\quad i,j=1,2 \,, \nonumber\\
\{Q_{i\a}, Q_{j\b}\}&=\{\bar Q^i{}_{\dot\a}, \bar Q^j{}_{\dot\b}\}=0 \,.
\label{d4algebra}
\end{align}
The coordinate transformation (\ref{thetatr}, \ref{theta12}, \ref{trafo2})
corresponds to a redefinition of the four-dimensional $\N=2$ supersymmetry
generators such that \mbox{$\exp(\theta_iQ^i + {\bar\theta}_i{\bar Q}^i) =
  \exp(\theta Q + \thetasl\Qsl$} $+ {\bar \theta} {\bar Q} + {\bar
  \thetasl}{\bar \Qsl})$ .  We define the new supersymmetry generators by
\begin{align} \label{generators}
  Q_\a \equiv \tilde Q_{1 \a} + i \tilde Q_{2 \a}, \quad \Qsl{}_{\a} \equiv
  \tilde \Qsl{}_{1 \a} + i \tilde \Qsl{}_{2 \a} \, , 
\end{align}
where
\begin{align}
  \tilde Q_{i\a} \equiv \textstyle\frac{1}{2} (Q_{i\a} + \bar
  Q^{i}{}_{\dot\a}), \quad \tilde \Qsl{}_{i\a} \equiv \textstyle\frac{i}{2}
  (Q_{i\a} - \bar Q^i{}_{\dot\a}), \quad i=1,2\,.
\end{align}
In terms of these new generators the algebra acquires the form
\begin{align} 
  \{  Q_\a, \bar{Q}_{\b} \} &= 2  \s^M_{\a\b}
  P_M, \quad \{\Qsl{}_\a, \bar{\Qsl}{}_{ \b} \}
  = 2 \s^M_{\a\b} P_M, \quad M=0,1,3 \, , \nonumber\\
 \{  Q_\a, {Q}_\b \}&=\{ \bar Q_{\a}, \bar{Q}_{ \b} \}=
 \{  \Qsl_\a, {\Qsl}_\b \}=\{ \bar \Qsl_{ \a}, \bar{\Qsl}_{ \b} \}=0\,,
\label{algebra}\\
\{ Q_\a, {\Qsl}{}_{\b} \} &= 
 \{ \Qsl_\a, {Q}{}_{\b} \}=0 \,,\quad
 \{ Q_\a, \bar{\Qsl}{}_{\b} \} = 
 \{ \Qsl_\a, \bar{Q}{}_{\b} \}=
- 2i \s^2_{\a\b} P_2 \,. 
\nonumber
\end{align}
Here we have made use of the fact that the Pauli matrices $\s^M$ are symmetric
while $\s^2$ is antisymmetric.  The algebra now splits into two $\N=2$, $d=3$ 
superalgebras, one generated by $Q_\a$, the other by $\Qsl_\a$. Both
superalgebras are connected via the generator $P_2$.\footnote{A related
  discussion of this algebra may be found in \cite{Lechtenfeld}.}

The corresponding superspace covariant derivatives, 
which anticommute with the supersymmetry generators (\ref{generators}) and
satisfy an algebra analogous to (\ref{algebra}), are given by
\begin{align} \label{Ds}
D&= \frac{\partial}{\partial \theta}
+ i \s^M \bar \theta \partial_M +  \s^2 \bar \thetasl \partial_2 \,,\quad
\bar D= - \frac{\partial}{\partial \bar \theta}
-i \theta \s^M \partial_M + \thetasl \s^2 \partial_2 \,,\\
\Dsl &=  \frac{\partial}{\partial \thetasl} + i \s^M \bar \thetasl
\partial_M - \s^2 \bar \theta \partial_2 \,, \quad
\bar \Dsl =  -\frac{\partial}{\partial \bar \thetasl} - i \thetasl \s^M 
\partial_M - \theta \s^2 \partial_2 \,.
\end{align}

The subspace defined by $\thetasl = 0$ is preserved by the $\N=2$, $d=3$
algebra generated by $Q$ and $\bar Q$.  If one introduces a superspace
boundary at
\begin{align}
x_2 = 0 \nonumber, \quad
\thetasl =0
\end{align}
the generators $P_2$, $\Qsl$ and $\bar \Qsl$ are broken,  leaving the unbroken
$\N=2$, $d=3$ supersymmetry algebra  
\begin{align} \label{alg3}
  \{ Q_\a, \bar{Q}_{\b} \} &= 2 \s^M_{\a\b} P_M \,,\nonumber\\ 
\,\quad\{  Q_\a, {Q}_\b \}&=\{ \bar Q_{\a}, \bar{Q}_{ \b} \}=0\,.
\end{align}
At the boundary, the derivatives $D$ and $\bar D$ given by (\ref{Ds}) reduce
to three-dimensional $\N=2$ covariant derivatives
\begin{align}
{D}&= \frac{\partial}{\partial \theta}
+ i \s^M \bar \theta \partial_M  \,,\quad
{\bar D}= - \frac{\partial}{\partial \bar \theta}
-i \theta \s^M \partial_M  
\end{align}
which satisfy the $\N=2$, $d=3$ algebra
\begin{align}
\{  {D}_\a, \bar{{D}}_{ \b} \} &
  = -2 \s^M_{\a\b} P_M, \quad M=0,1,3\quad, \nonumber\\
\{  {D}_\a, {{D}}_{\b} \}&=
\{  \bar{D}_{\a}, \bar{{D}}_{\b} \}=0 \,.
\end{align}

\subsection{Decomposition of the 4d vector multiplet under 3d $\N=2$} 
\label{sec2.2}
We now decompose the four-dimensional $\N=2$ abelian vector superfield $\Psi$
into 3d $\N=2$ superfields by performing the transformation (\ref{thetatr})
and subsequently setting $\thetasl=0$.  We show that the $\N=2$, $d=4$ vector
supermultiplet $\Psi$ decomposes into
\begin{align} \label{decomposition}
  \Psi\vert_{\thetasl=0}= \,\frac{1}{2} (\Phi + \bar\Phi + \sqrt{2}
  \,\partial_2V) + i \frac{1}{\sqrt{2}}\Sigma \,,
\end{align} 
where $\Phi$, $\bar \Phi$ are chiral and antichiral $\N=2$, $d=3$ 
supermultiplets, respectively. The 3d $\N=2$ linear supermultiplet $\Sigma$
is related to the 3d vector superfield $V$ by
\begin{align} 
  \Sigma(x,\theta,\bar\theta)\equiv\frac{1}{2i} \ve^{{\a}\b} \bar D_{ \a} D_\b
  V(x,\theta,\bar\theta) \,, \label{Sigma}
\end{align}
and satisfies $DD\Sigma= \bar D \bar D \Sigma=0$.  Note that in the definition
of an abelian linear multiplet the order of the derivatives is unimportant
since $\ve^{{\a}\b} \bar D_{ \a} D_\b V(x,\theta,\bar\theta) = \ve^{{\b\a}}
D_\b \bar D_{ \a} V(x,\theta,\bar\theta)$.

We start from the decomposition of the vector multiplet $\Psi$ under 
$\N=1$, $d=4$ which is given by an expansion in $\theta_2$ \cite{Lykken},
\begin{align} \label{N=1expansion}
  \Psi(\tilde y,\theta_1,\theta_2)=\Phi'(\tilde y,\theta_1) + i \sqrt{2}
  \theta^\a_2 W'_\a(\tilde y,\theta_1) + \theta_2 \theta_2 G'(\tilde
  y,\theta_1)\,\,,
\end{align}
where the $\N=1$ chiral and vector multiplets $\Phi'$ and $W'_\a$ have the
expansions
\begin{align}
  \Phi'(\tilde y,\theta_1) =&\, \phi'(\tilde y) + \sqrt{2}\theta_1 \psi'(\tilde y)
  + \theta_1\theta_1 F'(\tilde y) \,, \nonumber\\
  W'_\a(\tilde y,\theta_1) =&-i\l'_\a(\tilde y) + \theta_{1\a} D'(\tilde y) -
  \frac{i}{2}(\s^\m \bar \s^\n \theta_1)_\a F'_{\m\n}(\tilde y)
  + (\theta_1\theta_1) \s^m_{\a\dot\b} \partial_m \bar \l'{}^{\dot \b}(\tilde y)
 \,.
\end{align}
While the bosonic components $\phi'$ of $\Phi'$ and $v'_\mu$ of $W'_\a$ are
singlets under the global $SU(2)$ R symmetry, the fermions $\psi'$ of $\Phi'$
and $\l'$ of $W'_\a$ form a $SU(2)$ doublet.  

We are now interested in the form of $\Psi$ as given by (\ref{N=1expansion})
in the coordinates 
\mbox{($\theta$, $\bar \theta$, $\thetasl$, $\bar \thetasl$)} with
$\thetasl=0$. Since the 4d $\N=2$ algebra in the form (\ref{algebra}) reduces
to the 3d $\N=2$ algebra (\ref{alg3}), we expect $\Psi$ to decompose into 3d
$\N=2$ superfields at $\thetasl=0$.  Taking the inverse of the coordinate
transformation (\ref{theta12}, \ref{trafo2}) and setting $\thetasl=0$, we get
\begin{align} \label{inversetrafo2}
\theta_1&= \frac{1}{2} (\theta + \bar \theta) \,,\quad
\theta_2=  \frac{i}{2} (\theta - \bar \theta) \,. \quad 
\end{align}
After substituting the coordinate transformation (\ref{inversetrafo2}) into
(\ref{N=1expansion}), we can rearrange the components of
$\Psi\vert_{\thetasl=0}$ into 3d $\N=2$ chiral $(\phi,\psi;F)$ and linear
multiplets $(\rho,v_M,\l;D)$.  For this purpose we define new scalars and
vectors by
\begin{align}
{\rm Re}\, \phi &\equiv {\rm Re}\, \phi'\,, \quad\label{redef1}
{\rm Im}\, \phi \equiv \frac{1}{{\sqrt{2}}}\, v'_2 \,,\quad
\rho \equiv {\sqrt{2}}\, {\rm Im}\, \phi'\,, \\
F_{MN} &\equiv F'_{MN}\,,
\end{align}
and also new complex spinors by
\begin{align}
\psi &\equiv  {\rm Re\,}\psi' + i\, {\rm Re\,}\l'\,, \quad 
\bar \psi \equiv  {\rm Re\,}\psi' - i \,{\rm Re\,}\l'\,,\nonumber\\
\l &\equiv  {\rm Im\,}\psi' - i\, {\rm Im\,}\l',\quad
\bar \l \equiv  {\rm Im\,}\psi' + i \,{\rm Im\,}\l'\,.
\end{align}
Here we combined a 3d $\N=1$ scalar vector multiplet $({\rm Im}\, \phi', {\rm
  Im}\,\psi'; F')$ and a vector multiplet $(v'_M, {\rm Im}\, \l')$ to a $\N=2$
vector multiplet \cite{Nishino} or, more precisely, to a $\N=2$ linear
multiplet $(\rho,\l, v_M; D)$. Let us also define new auxiliary fields
\begin{align}
F &\equiv \frac{1}{2}(F'-F'{}^*-\sqrt{2}D')  \,, \nonumber\\
D &\equiv \frac{1}{\sqrt{2}} (F'+F'{}^*) + \sqrt{2} \,\partial_2 {\rm Re\,}\phi' \,.
\label{redef2}
\end{align}  
The term $\partial_2 {\rm Re\,}\phi'$ in the definition of the auxiliary field
$D$ seems unnatural at first sight but is required by $\N=2$, $d=3$ 
supersymmetry. The transverse derivative $\partial_2$ appears due to the
expansion in the 2-direction since the bosonic coordinates $\tilde y^\m$
differ from $x^\m$ only in the transverse direction, i.e.\ if $\m=2$,
\begin{align}
  \tilde y^\m \equiv x^\m +i \theta_1 \s^\m \bar \theta^1 + i \theta_2 \s^\m
  \bar \theta^2= x^\m + i \theta \s^2 \bar \theta \d^\m_2\,.
\end{align}
With the above definitions $\Psi\vert_{\thetasl=0}$ can be expressed
completely in terms of 3d $\N=2$ superfields. A detailed calculation in
App.\ \ref{App1b} shows that $\Psi\vert_{\thetasl=0}$ can be written as
\begin{align} \label{decomposition2}
  \Psi\vert_{\thetasl=0}= \,\frac{1}{2} (\Phi + \bar \Phi +\sqrt{2}
  \,\partial_2V) + i \frac{1}{\sqrt{2}}\Sigma \,,
\end{align} 
where $\Sigma$ and $\Phi + \bar\Phi$ read in components 
\begin{align}
 \Sigma(x,\theta,\bar\theta)=&\,\rho+ \theta \bar\l+\bar\theta\l +i
  \theta \bar\theta D + \frac{1}{2} \bar\theta \s_{K} \theta
  \varepsilon^{MNK} F_{MN} \nonumber\\
  &+ \frac{i}{2} (\theta\theta) \bar\theta \bar \s^M 
  \partial_M \bar \l + \frac{i}{2} (\bar\theta\bar\theta) \theta \s^M 
  \partial_M \l - \frac{1}{4} (\theta\theta)(\bar\theta\bar\theta)
  \square_3 \rho \,, \\
\Phi + \bar\Phi =&\, (\phi + \phi^*) + \sqrt{2} \theta \psi
+ \sqrt{2} \bar \theta \bar \psi +\theta\theta F
+ \bar\theta \bar\theta F^* + i \theta \s^M \bar\theta \partial_M
(\phi - \phi^*) \nonumber\\
&+ \frac{i}{\sqrt{2}} (\theta\theta) \bar \theta \bar \s^M \partial_M \psi
+ \frac{i}{\sqrt{2}} (\bar\theta\bar\theta) \theta \s^M \partial_M \bar \psi
-\frac{1}{4}(\theta\theta)(\bar\theta\bar\theta)\square_3(\phi + \phi^*)\,.
\end{align}
The component expansion of the linear superfield $\Sigma$ is derived in App.\ 
\ref{App1a}.  Eq.\ (\ref{decomposition2}) coincides with
(\ref{decomposition}).

\section{A superconformal $\N=2$, $d=4$ theory with conformal
boundary couplings} \setcounter{equation}{0}

\subsection{$\N=2$ d=4 action with manifest $\N=2$ d=3 supersymmetry}

With the help of the decomposition (\ref{decomposition}) it is now
straightforward to construct the action for a $\N=2$, $d=4$  vector
supermultiplet in a $\N=2$, $d=3$  superspace. As discussed in
section 2, the degrees of freedom of the four-dimensional $\N=2$
vector multiplet $\Psi$ are contained in a  $\N=2$, $d=3$ vector
multiplet $V$ and a $\N=2$, $d=3$ chiral multiplet $\Phi$. Strictly
speaking, there are continuous sets of such multiplets labelled by
the coordinate $z\equiv x_2$ transverse to the boundary or defect.
It will be convenient to work with the linear multiplet $\Sigma $,
which is related to $V$ by (\ref{Sigma}).  Written in terms of
$V,\Sigma$ and $\Phi$, the action of the free Abelian $\N=2$, $d=4$
vector multiplet becomes
\begin{align}
  S^{\rm 4d}_{\rm bulk}&= \frac{1}{8\pi} {\rm Im} \left[ \tau \int \di z
    \di^3x \di^2\theta \di^2{\bar\theta}\, \left(\sqrt{2} \Sigma + i 
(\sqrt{2}
      \partial_z V + {\bar \Phi} + \Phi ) \right)^2
  \right]\label{action1} \nonumber\\
  &= \frac{1}{g^2} \int \di z \di^3x \di^2\theta \di^2{\bar\theta} \, \left[
    \Sigma^2 - \frac{1}{2}(\sqrt{2}\partial_z V
    + {\bar \Phi} + \Phi)^2 \right]\\
  &\qquad\qquad\quad+ \frac{\theta_{\rm YM}}{16\pi^2} \int \di z \di^3x
  \di^2\theta \di^2{\bar\theta}\, \sqrt{2} \Sigma(\sqrt{2} \partial_z V +
  {\bar \Phi} + \Phi)\,, \nonumber
\end{align}
where $\tau = \frac{\theta_{\rm YM}}{2\pi} + \frac{4\pi i}{g^2}$.
In the case of a boundary at $z=0$, the $z$ integration runs from
$0$ to $\infty$, whereas for a defect the $z$ integration runs
from $-\infty$ to $\infty$.  The bulk action (\ref{action1}) has
manifest three-dimensional Lorentz invariance.  Four-dimensional
Lorentz invariance is not manifest, and is explicitly broken by
the introduction of a boundary or defect. In the absence of
either, four-dimensional Lorentz invariance can be seen in
component notation after integrating out auxiliary fields. For
instance the kinetic terms in the $z$ direction such as
$\partial_z \phi^{\prime *}\partial_z \phi^{\prime}$ arise upon
integrating out the auxiliary $D$ term.

Note that the term proportional to the theta angle in
(\ref{action1}) is a total derivative in four dimensions, which
can ordinarily be ignored in an abelian theory.  However in the
presence of a three-dimensional boundary at $z=0$, it can be
rewritten as a boundary Chern-Simons term of the form
\begin{align}
  S_{CS} = \frac{\theta_{\rm YM}}{8 \pi^2} \int_{z=0} \di^3 x \di^2\theta
\di^2{\bar \theta}\,\Sigma V \, .
\end{align}
The terms involving the products $\Sigma\Phi$, $\Sigma \bar \Phi$
in (\ref{action1}) vanish after integrating the derivatives
contained in $\Sigma$ by parts.

The action (\ref{action1}) is invariant under four-dimensional
gauge transformations given by
\begin{align}
V \rightarrow  V + \Lambda + {\bar \Lambda},\quad \Sigma
\rightarrow \Sigma,\quad \Phi\rightarrow \Phi - \sqrt{2}
\partial_z\Lambda \,.
\end{align}
where $\Lambda(\theta, \vec{x},  z)$ are $\N=2$, $d=3$ chiral superfields,
labelled by the continuous index $z$.

\subsection{Boundary Interaction}

We now couple the bulk action (\ref{action1}) to a
three-dimensional theory living on a defect or boundary at
$z\equiv x_2=0$. In the following discussion we consider the boundary
case. Our results concerning the action and its renormalization
properties are also valid for the defect case  since they do not
depend on the imposition of boundary conditions at least in the
abelian case  considered here.

We may choose either Dirichlet or Neumann boundary conditions. In
$\N=2$, $d=3$ superspace, Dirichlet boundary conditions are given
by
\begin{gather} \label{Dirichlet}
\Sigma\vert_{z=0}=0 \, .
\end{gather}
which implies $F_{MN} =0$ at the boundary.  We shall instead
choose Neumann boundary conditions given by
\begin{gather} \label{Neumann}
(\sqrt{2} \pr_z V + \Phi + \bar \Phi)\vert_{z=0} = 0
\end{gather}
implying $F_{M2} =0$ at $z=0$.  This choice is suitable for
introducing couplings to electrically charged matter at the
boundary.

The boundary breaks half the bulk supersymmetries,
leaving only $\N=2$, $d=3$ invariance. We shall couple half the
bulk degrees of freedom, i.e. the $\N =2$, $d=3$ vector multiplet
$V$,  to charged $\N=2$, $d=3$ chiral multiplets living at the
boundary.
The action consists of two parts,
\begin{equation}
S =S^{\rm 4d}_{\rm bulk}+ S^{\rm 3d}_{\rm boundary} \, .
\end{equation}
For the bulk action we take free abelian $\N=2$, $d=4$ theory as
given in $\N=2$, $d=3$ superspace (\ref{action1}),
\begin{gather}
S^{\rm 4d}_{\rm bulk} = \frac{1}{g^2} \, \int \di z \di^3 x
\di^2\theta \di^2{\bar \theta} \left[ \Sigma^2 -
\frac{1}{2}(\sqrt{2} \partial_z V + \Phi + {\bar \Phi})^2\right]
\,.
\end{gather}
The boundary action includes both the boundary field kinetic term
and the interactions between bulk and boundary fields. For our
model we consider the boundary degrees of freedom to be given by
chiral superfields $B^+$ and $B^-$ of opposite charge. Under gauge
transformations
\begin{gather}
B^+ \rightarrow e^{i\Lambda} \Bp \,, \quad 
B^- \rightarrow e^{-i\Lambda} B^- \, , \quad {\rm
with} \; \Lambda \, = \, \Lambda(\theta, \vec{x}, z=0)\, .
\end{gather}
Together with a possible
Chern-Simons term, the boundary part of the action is
\begin{align} \label{boundary1}
S^{\rm 3d}_{\rm boundary} = & \,  \, \int\! \di^3x \di^2 \theta
\di^2\bar \theta \,  \bar B^{\pm} e^{\pm g V} B^{\pm} + \,
\frac{\theta_{\rm YM}}{8 \pi^2} \int \di^3 x\, \di^2 \theta
\di^2\bar\theta \,V \Sigma \, ,
\end{align}
where $^\pm$ denotes summation over $\Bp$ and $\Bm$.

The combined action  $S= S^{4d}_{\rm bulk} + S^{3d}_{\rm
boundary}$ is classically invariant under conformal symmetries
which leave the boundary invariant. We note that classically
the three-dimensional R weights under the the $U(1)_R$ group which
determines the supercurrent multiplet are $R(B^\pm)=\frac{1}{2}$,
$R(\bar B^{\pm}) = - \frac{1}{2}$, $R(\Phi)=1$, $R(\bar \Phi)=-1$
and $R(V)=0$. The classical dimensions are given by $D(B^{\pm }) =
\frac{1}{2}$, $D(\Phi) =1$ and $D(V) = 0$.  The dimensions in a
superconformal $\N=2,d=3$ theory satisfy the inequality $D \geq
|R|$ \cite{Aharony} which must be saturated for the chiral
primaries $B^{\pm}$ and $\Phi$.

\subsection{Superconformal transformations and Renormalization}
We proceed by studying the renormalization properties of our theory.  It is
crucial to note that it suffices to consider the renormalization of the
boundary 1PI action corresponding to (\ref{boundary1}) in view of obtaining
the $\beta$ functions since all vertices are three-dimensional and since our
theory is abelian. The $d=4$ part of the 1PI action is finite by construction.
Nonetheless the boundary action potentially receives quantum corrections from
propagation through the bulk. We derive a Callan-Symanzik equation for the
boundary theory by studying its superconformal transformation properties.

We obtain the superconformal transformations of the fields by adapting results
from $\N=1$, $d=4$ theory \cite{PS,ERS}. The generator of $\N=2$, $d=3$
superconformal transformations is given by
\begin{align} \label{W}
  W = \int \! \di^3x \di^2 \theta \di^2 \bar \theta \, \big[ & \Omega^\alpha
  \big( w_\alpha (B^{\pm}) + w_\alpha (V)\, \big) \, + \bar \Omega_\beta \big(
  \bar w^\beta (\bar B^{\pm}) + \bar w^\beta (V)\, \big) \big] \, .
\end{align}
Here $\Omega, \bar \Omega$ are the parameters of the 
superconformal transformations which satisfy $ D^\alpha \bar
\Omega^\beta = \bar D^\beta \Omega^\al$. For the local superconformal 
transformations of the fields we have
\begin{align}
w_\alpha(B^\pm) & =  {\ \frac{1}{4}} \big( D_\alpha B^{\! \pm} 
\frac{\delta}{\delta B^{\! \pm}} - \frac{1}{4}\, D_\al ( B^{\! \pm}
\frac{\delta}{\delta \Bpm} ) \big) \, , \nonumber\\
w_\alpha(V) & =  \frac{1}{2} \bar D^\beta ( \bar D_\beta D_\alpha V
\frac{\delta}{\delta V} ) + \frac{1}{4} \bar D^2 ( D_\al V
\frac{\delta} {\delta V})  \, , \label{localconf}
\end{align}
where the factor of $\frac{1}{4}= \frac{1}{2}R_B$ in the expression for
$w_\alpha(B^\pm)$  is determined by the R weight $R_B=\frac{1}{2}$ of $\Bpm$. 
We note that the classical theory given by
 (\ref{boundary1}) is superconformally invariant, $WS=0$.
Applying (\ref{localconf}) to the action (\ref{boundary1}) gives
\begin{gather} \label{walpha}
w_\al S \,\equiv \, (w_\al(\Bpm) + w_\al(V))\,
S  = \, \bar
D^\beta {\cal J}_{\al \beta}  \, ,
\end{gather}
with ${\cal J}_{\al \beta}$ the supercurrent multiplet. 
Upon quantization there will be a potential trace anomaly $D_\al {\cal
T}$, with ${\cal T}$ chiral, contributing to the r.h.s.~of
(\ref{walpha}), whose explicit form is discussed in detail below. 

For scale transformations and for R transformations we have 
\begin{align} \label{DD}
\Omega^\alpha{}_D &= 
\frac{1}{2}\theta^\alpha - \frac{i}{2} x^M \sigma_M^{\alpha \beta}
\bar \theta_\beta \,, \\ \Omega^\alpha{}_R & = \, i \theta^\al \bar
\theta^2
\, ,
\end{align}
respectively, for which (\ref{W}) defines the transformation operators
$W^D$ and $W^R$. From dimensional analysis the 1PI action satisfies
\begin{gather} \label{WD}
( \mu \frac{\partial}{\partial \mu} \, + \, W^D ) \, \Gamma^{\rm 3d}_{\rm bdy}  \, = 0 \, , 
\end{gather}
with $\mu$ the renormalization scale. 

For investigating the superconformal transformation properties of the
quantized theo\-ry in a perturbation expansion to all orders, we have
to ensure well-defined finite local operator insertions. For this
purpose we follow the BPHZ approach \cite{BPHZ}. 
This is very convenient in the
present situation since  our argument is based on
symmetry considerations for operator
insertions and we do not need to perform explicit  calculations
beyond one loop. Since the theory given by (\ref{boundary1}) is
massless, it requires regularization by an auxiliary mass term
which may be taken to zero at the very end of the calculation as
described below. With regularization, the BPHZ effective action
corresponding to (\ref{boundary1}) has the form
\begin{align}
\Gamma^{\rm 3d, \, eff}_{\rm boundary} = & \phantom{+}
 z_B \,  \int\!\di^3 x \di^2 \theta \di^2
\bar \theta \, \bar{B}^{\!\pm} e^{\pm g V} B^{\!\pm} \, 
\;  + \, z_v  \,  \int\!\di^3 x \di^2 \theta \di^2
\bar \theta \, V \Sigma \nonumber\\ & 
- \, M\, \Big( \int \! \di^3x \di^2 \theta \, B^+ B^- \, + \, 
 \int \! \di^3x \di^2 \bar \theta \, \bar B^+ \bar B^-  \Big)
. \label{boundaryM}
\end{align}
The BPHZ effective action is not to be confused with the Wilsonian or
1PI effective action and has the advantage of being local. It is
related to the non-local 1PI action via the {\it action principle}.
This means that for the derivative of the 1PI action with respect to a
field or coupling we have
\begin{gather} \label{insertion}
\frac{\delta}{\delta V} \, \Gamma^{\rm 1PI} \, = \,  \big[ \frac{ \;
\delta \Gamma^{\rm eff}}{\delta V} \big] \cdot \Gamma^{\rm 1PI} \, .
\end{gather}
Here the square brackets denote a well-defined finite local operator
insertion. The r.h.s.~of this equation is the generating functional
for 1PI Green functions with an insertion of the local operator
$\delta \Gamma^{\rm eff}/ \delta V$.

The field renormalization 
coefficients $z_B$ and $z_V$ in (\ref{boundaryM}) are perturbative 
power series in the coupling, starting with the classical value
\begin{gather}
z_B \, = \, 1 + \dots \, , \quad z_V \, = \, \theta_{\rm YM}\,  + \dots \, .
\end{gather}
Gauge fixing terms contributing to the 1PI action are also required in
principle. A possible gauge condition is given for instance by
$\bar D\bar D DD V + \pr_z \Phi =0$. However the gauge fixing terms are
not essential for the analysis of symmetry transformations performed
here, since their operator insertions vanish when acting on physical
states, ie.~inside Green functions.

For the superconformal transformation of the boundary 1PI action,
given by applying (\ref{W}) with (\ref{localconf}) to the 1PI action
corresponding to (\ref{boundaryM}),
we obtain
\begin{gather} \label{conftrans}
\int \! \di^3 x \di^2\theta\di^2\bar \theta \Big( \Omega^\al  \big( 
w_\alpha (B^{\pm})  + w_\alpha (V)\, \big)\, +\,
\bar \Omega_\beta  \big( 
\bar w^\beta (B^{\pm})  + \bar w^\beta (V)\, \big)
 \Big) \,    \Gamma^{\rm 3d}_{\rm bdy}  \hspace{2.6cm} 
\nonumber\\ \hspace{0.8cm} 
= \, -\frac{1}{8} \, \int\!
\di^3 x \di^2\theta\di^2\bar \theta \, \Big(
\Omega^\al D_\al  \, [ M \Bp \Bm ] \cdot \Gamma^{\rm 3d}_{\rm bdy}
+ \bar \Omega_\beta \bar D^\beta  \, [ M \bar \Bp \bar\Bm ] \cdot \Gamma^{\rm 3d}_{\rm bdy}
 \Big) \, , 
\end{gather}
with $[ M B^+ B^- ]$ a well-defined local mass insertion as defined in
(\ref{insertion}). This mass insertion potentially gives rise to a
chiral trace anomaly ${\cal T}$. According to
the standard BPHZ procedure we have to expand the mass insertion into
a {\it Zimmermann identity} \cite{Zimmermann}, 
reminescent of the operator product expansion,
before being able to set $M$ to zero. This gives
\begin{gather}
[M \Bp \Bm ] = [ u \bar D^2 \bar \Bpm e^{\pm gV} \Bpm + v \bar D^2 V
\Sigma + w (B^+ B^-)^2 ] + M [\Bp \Bm] \, ,\label{Zimmermann}
\end{gather}
with a similar relation for $[M \bar \Bp \bar\Bm]$ which is obtained
by complex conjugation.
The first bracket on the r.h.s.~contains a basis of local field
polynomials of the same dimension and chirality as the l.h.s., with
coefficients
$u$, $v$ $w$  of order ${\cal O} (\hbar)$ and higher. On the
r.h.s., $M$ may now safely be put to zero since the last term is a
so-called `soft' insertion. One of the key points in
view of the renormalization properties of the theory is now that
the coefficient $v$ vanishes due to gauge symmetry
requirements\footnote{For purely three-dimensional Chern-Simons
theories a similar non-renormalization argument for the Chern-Simons
term may be found in \cite{Piguet}.}:
The contribution to (\ref{Zimmermann}) involving the local Chern-Simons term
has to be absent since this term is not gauge invariant. 
Moreover the coefficient $w$ is zero due to the chirality of
$(\Bp\Bm)^2$, such that $R$ symmetry is preserved.
We note, however, that the coefficient $u$ in (\ref{Zimmermann}) is
non-zero in general.

With the help of the superconformal Ward identity we now derive a
Callan-Symanzik equation which will allow us to prove conformal
invariance for our model. The superconformal transformation of the 1PI
action is given by 
\begin{gather} 
\int\!\di^3x \di^2 \theta \di^2 \bar \theta \, \big( \Omega^\al 
w_\al + \bar \Omega_\beta 
\bar w^\beta \, \big)\, \Gamma^{\rm 3d}_{\rm bdy}
 \hspace{8cm}  \nonumber\\
 = \, -\frac{1}{8} \,
\int\!\di^3x \di^2 \theta \di^2 \bar \theta \, \big( \Omega^\al D_\al  
\bar D^2 \,  
+ \, \bar \Omega_\beta {\bar D}^\beta  D^2 \, \big) \,  
\left[  u \bar B^{\pm}
e^{\pm g V} B^{\pm} \right]\,  \cdot \Gamma^{\rm 3d}_{\rm bdy}
 \, , \label{superconftr}
\end{gather}
with $w_\alpha$ as in (\ref{localconf}), (\ref{walpha}). 
For scale transformations as
given by (\ref{DD}), (\ref{superconftr}) and (\ref{WD}) imply 
\begin{align}
\mu \frac{\pr}{\pr \mu} \Gamma^{\rm 3d}_{\rm bdy}
 = & -\frac{1}{2} \, \int\di^3x \di^2
\theta \di^2 \bar \theta \, u \, [ \bar B^{\pm}
e^{\pm g V} B^{\pm} ]  \cdot \Gamma^{\rm 3d}_{\rm bdy} \, . \label{scale1} 
\end{align}
Using (\ref{scale1}) as well as the action principle (\ref{insertion})
and the Zimmermann identity (\ref{Zimmermann}),
we derive the Callan-Symanzik equation by making
use of the fact that derivatives with respect to the fields and
couplings give rise to local insertions of the form
\begin{gather} \label{cderiv}
\gamma_B \Bpm \frac{\delta}{\delta \Bpm} \, \Gamma^{\rm 3d}_{\rm bdy}
 \, = \,  (\gamma_B 
z_B - 2 \gamma_B u) \,  [{\bar D}{\bar D} \bar B^{\pm}
e^{\pm g V} B^{\pm}] \cdot \Gamma^{\rm 3d}_{\rm bdy} \, ,
\end{gather}
and similar results for $V$ and $g$. 
We obtain the Callan-Symanzik equation
by comparing the coefficients of the insertions.
In the present case the Callan-Symanzik equation has just the simple form
\begin{gather} 
( \mu \frac{\pr}{\pr\mu}  -  \gamma_B \N_B ) \, \Gamma^{\rm 3d}_{\rm bdy}
 = 0 \, , \label{CSE}\\ \qquad \N_B \equiv \int\!\di^3x \di^2 \theta \, B^\pm
 \frac{\delta}{\delta \Bpm} +
\int\!\di^3 x \di^2 \bar \theta \, \bar B^\pm
 \frac{\delta}{\delta \bar \Bpm} \, .\nonumber
\end{gather} 
Subject to  the condition
\begin{gather}
 4 \gamma_B z_B + (1-8 \gamma_B) u  =  0 \, , 
\end{gather}
(\ref{CSE}) coincides with (\ref{scale1}). The term involving $u$ in
(\ref{scale1}) has been absorbed into an anomalous dimension for the
chiral boundary fields. This anomalous dimension is non-zero, as we
confirm by an explicit one-loop calculation in appendix B.3. 
The beta function in the Callan-Symanzik equation vanishes, such that
we have a conformal theory. We may also write a superconformal Ward
identity expressing superconformal invariance. Using
\begin{align}
w^{(\gamma)}_\al & \equiv  w^{(\gamma)}_\al (\Bpm) + w_\al (V) \, ,
\nonumber\\
w^{(\gamma)}_\al (\Bpm) & = \frac{1}{4} \left(   D_\al \Bpm
\frac{\delta}{\delta \Bpm} - \frac{1}{4}(1 + 2 \gamma_B) D_\al (\Bpm
\frac{\delta}{\delta \Bpm} ) \right) \, , 
\end{align}
we have 
\begin{align}
\int \! \di^3x\di^2 \theta \di^2 \bar \theta \, \left( 
\Omega^\al
w^{(\gamma)}_\al + \bar \Omega_\beta
\bar w^{(\gamma)}{}^\beta \right) \, \Gamma^{\rm 3d}_{\rm bdy} & = 0 \,.
\end{align}
This shows explicitly that the
 theory is superconformal with the boundary fields
acquiring anomalous dimensions.

\newpage

\section{An abelian $\N=4$ SCFT with boundary}\setcounter{equation}{0} 
\subsection{$\N=4$ d=4 action with manifest $\N=2$ d=3 supersymmetry} 
We now turn to the case of $\N=4$ supersymmetry. In this case we have to
consider an \mbox{$\N=2$} hypermultiplet in the bulk in addition to the $\N=2$
vector multiplet considered before. Similarly to the decomposition of the
$\N=2$, $d=4$ vector multiplet under $\N=2$, $d=3$ in section \ref{sec2.2},
the degrees of freedom of the hypermultiplet can be rearranged into two
$\N=2$, $d=3$ chiral superfields $Q_1$ and $Q_2$.  The bulk multiplets
($\Sigma, Q_2$) and ($\Phi, Q_1$) fit into $\N=4$, $d=3$ linear
and hypermultiplets with the  bosonic and fermionic components
\begin{align}
  \Sigma, Q_2\,\, &\rightarrow \quad \rho, q_2 \in (3,1),\quad \l, \l_2 \in 
  (2,2), v_M \in (1,1)\,,\nonumber\\
  \Phi, Q_1\,\, &\rightarrow \quad {\rm Re}\,\phi, q_1 \in (1,3),\quad
   \psi, \l_1 \in (2,2), \quad {\rm Im}\,\phi \in (1,1)\,,
\end{align}
where $(r_V,r_H)$ denotes the representation of $SU(2)_V \times SU(2)_H
\subset SU(4)$. The components of $\Sigma$ and $\phi$ are given by the
analysis of section 2. 
The multiplet $Q_2$ contains the complex scalar $q_2$ which
we have chosen to be $q_2 \equiv {\rm Im\,} \phi'_1 - i {\rm Im\,} 
\phi'_2$ with
$\phi'_1$ and $\phi'_2$ being the scalars of the 4d hypermultiplet.  In $\N=2$
notation, the $SU(2)_V$ symmetry of the triplet 
$(\rho, {\rm Re}\,q_2, {\rm Im}\,q_2)$ will
not be manifest in the action but is required by the \mbox{R symmetry} of
$\N=4$, $d=3$. In our conventions the six scalars of $\N=4$
supersymmetry are given by $X_V=({\rm Im} \phi', {\rm Im} \phi'_1,
{\rm Im} \phi'_2)$ and $ X_H=({\rm Re} \phi', {\rm Re} \phi'_1,
{\rm Re} \phi'_2)$, with $\rho = \sqrt{2} \, {\rm Im} \phi'$ as in (2.18).  
 
The bulk action is obtained by rewriting the standard $\N=4$, $d=4$ SYM action
in $\N=2$, $d=3$ language.  In terms of superfields $\Sigma, \Phi, Q_1$, and
$Q_2$, we find
\begin{align}\label{bulkM2}
S_{\rm bulk} =
\frac{1}{g^2} \int \di z \di^3x \di^2{\theta} \di^2{\bar\theta}
\left(
\Sigma^2 - \frac{1}{2}(\sqrt{2} \partial_z V + \Phi + {\bar\Phi})^2
+ {\bar Q}_i Q_i \right) \nonumber \\
+ \int \di z \di^3x \di^2{\theta} \,
\epsilon_{ij}Q_i\partial_zQ_j + \int \di z \di^3x \di^2{\bar\theta}\, 
\epsilon_{ij}{\bar Q}_i\partial_z{\bar Q}_j \,.
\end{align}
The first term is the same as in our first model, cf.\ Eq.\ (\ref{action1}).
The remaining terms are kinetic terms for $Q_1$ and $Q_2$.
The chiral part $\epsilon_{ij}Q_i\partial_z Q_j$ is the four-dimensional
Lorentz completion of the ${\bar Q}_iQ_i$ term, as can be seen
in component notation after integrating out auxiliary fields.
Under ${\cal N}=4$, $d=3$ supersymmetry, $V$ and $Q_2$ belong to a
vector multiplet, while $\Phi$ and $Q_1$ belong to a hypermultiplet.

One may also include a Chern-Simons term at the boundary or defect. 
However there is no off-shell $\N=4,d=3$ 
Chern-Simons term \cite{Nishino,noCS2,Kapustin}.  
This may seem surprising since one might expect to be able to induce
such a term at a boundary by writing the bulk $\theta_{YM}$ term in 
$\N=4,d=3$ superspace.  However such a superspace representation need
not exist.  We may consider instead adding a Chern-Simons term
which breaks the the supersymmetry further to $\N=3, d=3$.  
The abelian $\N=3,d=3$ Chern-Simons term \cite{ZK,KLL} is written in 
$\N=2,d=3$ 
superspace as
\begin{align}
  S_{CS} = \frac{\theta_{\rm YM}}{8 \pi^2} \left[ \int \di^3x \di^2\theta
    \di^2{\bar \theta} \,\Sigma V + \int \di^3x \di^2{\theta}\, Q_2{}^2 + \int
    \di^3x \di^2{\bar\theta} \,{\bar Q}_2{}^2 \right] \,.
\label{CSthree}
\end{align}
Note that in a renormalizable model with $\N=3,d=3$ supersymmetry 
written in $\N=2,d=3$ superspace,  all the
terms except the Chern-Simons term 
preserve $\N=4,d=3$ supersymmetry (see \cite{Kapustin}).

\subsection{Boundary Interaction}

As above, we couple the 4d bulk action to a three-dimensional theory on the
defect or boundary at $z=0$, where the supersymmetry is broken
down to $\N=4$, $d=3$ such that only $SU(2)_V \times SU(2)_H \subset SU(4)$ is
preserved.  The boundary $\N=2$ superfields $B^+, B^-$ form a $\N=4$, $d=3$
hypermultiplet with bosonic components $b^+,b^- \in (1,2)$ and fermionic
$\chi^+,\chi^-\in (2,1)$. The $SU(2)_V$ symmetry of the doublet ($\chi^+,\bar
\chi^-$) will not be visible in the boundary action due to the $\N=2$,
$d=3$ language.

There are again two options for choosing boundary conditions.  We could impose
Dirichlet boundary conditions on the linear multiplet $(\Sigma, Q_2)$,
\begin{align}
\Sigma \vert_{z=0} = 0, \quad Q_2\vert_{z=0}=0\,,
\end{align}
and leave the hypermultiplet unconstrained.  This is not an adequate option as
there is no coupling to the boundary hypermultiplet. It is however possible to
work with a {\em twisted} hypermultiplet $(\hat B^+,\hat B^-)$ which is
related to an ordinary hypermultiplet by interchanging the group $SU(2)_V$
with $SU(2)_H$, i.e.\ the components $\hat b^+,\hat b^- \in (2,1)$ and
$\hat\chi^+,\hat\chi^-\in (1,2)$. We will not pursue this option.

Instead, we can extend the Neumann boundary conditions (\ref{Neumann}) to
\begin{align} \label{quasiNeumann}
\left( \partial_z V + \Phi + \bar \Phi \right) \vert_{z=0} = 0, 
\quad Q_1 \vert_{z=0}=0\,,
\end{align}
and couple the linear multiplet to the boundary.

With these boundary conditions the
 action, which consists again of two parts, is given by
\begin{equation} \label{abelianaction}
S \, = \, S^{\rm 4d}_{\rm bulk}+ S^{\rm 3d}_{\rm boundary} \, ,
\end{equation}
with $S^{\rm 4d}_{\rm bulk}$ given by
   (\ref{bulkM2}) and with the classical boundary action given by
\begin{align}\label{boundaryM21}
S{}^{\rm 3d }_{\rm bdy} =& 
 \int d^3 x d^2\theta d^2{\bar\theta} 
\left(
{\bar B}^+ e^{gV} B^+ + {\bar B}^- e^{-gV} B^- \right)
+ \frac{ig}{\sqrt{2}} \left[ \int d^3x d^2{\theta} B_+ Q_2 B_- +c.c. \right] \nonumber\\
& +  \,  \frac{\theta_{\rm YM}}{8 \pi^2} \left[\int\!\di^3 x \di^2 \theta \di^2
\bar \theta \, V \Sigma +  \,  \int\!\di^3x \di^2 \theta\, {Q_2}^2
+  \,  \int\!\di^3x \di^2 \bar \theta \, {\bar Q_2}{}^2 \right] \,.
\end{align}
In the first line we couple the vector multiplet $(V, Q_2)$ to
the charged boundary fields $B^+,B^-$. The terms involving
$Q_2{}^2$ and $\bar Q_2{}^2$ are the $\N=3$, $d=3$ supersymmetry 
completions of the Chern-Simons term $V\Sigma$.

\subsection{Renormalization}
Quantum conformal invariance of this $\N=4$ model was already demonstrated in
\cite{Freedman} using power counting and symmetry arguments in
component notation. 
Here we use again the BPHZ approach within $\N=2$, $d=3$ superspace
in order to prove the finiteness of the
theory. In addition to conformal invariance we also show that - unlike
in the $\N=2$ model of section 3 - the
elementary fields do not acquire anomalous dimensions.
Again it is sufficient to consider the boundary contribution to the
1PI action in view of determining the renormalization properties of
the complete model. 

For this purpose we add an auxiliary mass term to the boundary action
(\ref{boundaryM21}) for regularization and obtain for the BPHZ
effective action
\begin{align}\label{boundaryM2}
\Gamma{}^{\rm 3d , \, eff}_{\rm bdy} =& \; 
z_B \,  \int d^3 x d^2\theta d^2{\bar\theta} 
\left(
{\bar B}^+ e^{gV} B^+ + {\bar B}^- e^{-gV} B^- \right)
\nonumber\\ &
+ \frac{ig}{\sqrt{2}}  \left[ \int d^3x d^2{\theta} B_+ Q_2 B_- +c.c. \right] \nonumber\\
& + z_v  \,  \int\!\di^3 x \di^2 \theta \di^2
\bar \theta \, V \Sigma +  z_{Q_2} \,  \int\!\di^3x \di^2 \theta\, {Q_2}^2
+ z_{Q_2} \,  \int\!\di^3x \di^2 \bar \theta \, {\bar Q_2}{}^2 
\nonumber\\ & 
- \, M\, \Big( \int \! \di^3x \di^2 \theta \, B^+ B^- \, + \, 
 \int \! \di^3x \di^2 \bar \theta \, \bar B^+ \bar B^-  \Big)
\,.
\end{align}
The local superfield transformation of the 1PI action is now given by
\begin{align} 
\int\! \di^3x \di^2\theta & \di^2\bar \theta \, 
 \Big( \Omega^\al w_\al
+ \bar \Omega_\al \bar w^\al
 \Big)
 \Gamma^{\rm 3d}_{\rm bdy}  \nonumber\\ = &  
 \, -\frac{1}{8} \, \int\!
\di^3 x \di^2\theta\di^2\bar \theta \, \Big(
\Omega^\al D_\al  \, [ M  \Bp \Bm ] \cdot \Gamma^{\rm 3d}_{\rm bdy}
+ \bar \Omega_\beta \bar D^\beta   \, 
[ M \bar \Bp \bar\Bm ] \cdot \Gamma^{\rm 3d}_{\rm bdy} \,\Big)
 \, , \nonumber\\
w_\al \,  \equiv & \,  \label{410}
w_\alpha (B^{\pm}) + w_\alpha (Q_2) + w_\alpha (V)  \, ,
\end{align}
with the superconformal field transformations as in (\ref{localconf}) and 
\begin{align}
w_\alpha(Q_2) & = \, {\ \frac{1}{4}} \big( D_\alpha Q_2
\frac{\delta}{\delta Q_2} - \frac{1}{2}  \,  D_\al ( Q_2
\frac{\delta}{\delta Q_2} ) \big) \, . 
\end{align}
The {Zimmermann identity} as in (\ref{Zimmermann}) has now the
form
\begin{gather}
[M \Bp \Bm ] = [ u \bar D^2 \bar \Bpm e^{\pm gV} \Bpm  + v \Bp Q_2 \Bm
+ w \bar D^2 V
\Sigma  + y {Q_2}^2] + M [\Bp \Bm] \, .\label{Zimmermann2}
\end{gather}
The coefficient $w$ vanishes again by the gauge non-invariance of $\bar D^2 V
\Sigma$. $v$ and $y$  vanish due 
to the chirality of $B^+ Q_2 B^-$ and ${Q_2}^2 $, such that the
three-dimensional 1PI action is invariant under the
three-dimensional $U(1)_R$ symmetry transformation
generated by
\begin{align} \label{WR}
  W^R = \int \! \di^3x \di^2 \theta \di^2 \bar \theta \, \big[  \Omega^{R\alpha} &\big( w_\alpha
  (B^{\pm}) + w_\alpha (V) + w_\alpha (Q_2)\, \big) \, \nonumber \\
  & \qquad + \bar
  \Omega^R_{\beta} \big( \bar w^\beta (\bar B^{\pm}) + \bar w^\beta
  (V) 
+ \bar w^\beta
  (Q_2)\, \big) \big] \, , 
\end{align}
with $\Omega_\al{}^R = i \theta_\al \bar \theta^2$. 
Similarly chirality ensures the absence of any quartic $(\Bp\Bm)^2$
term.

After using the Zimmermann identity, we may safely set the soft mass term
to zero and obtain, with $ w_\al $ as in (\ref{410}),
\begin{gather} \label{superconftr2}
\int\!\di^3x \di^2 \theta \di^2 \bar \theta \, \big( \Omega^\al 
w_\al  + \bar \Omega_\beta 
\bar w^\beta \, \big)\, \Gamma^{\rm 3d}_{\rm bdy} \hspace{3cm} \nonumber\\ 
\hspace{1cm}  = \, -\frac{1}{8} \,
\int\!\di^3x \di^2 \theta \di^2 \bar \theta \, \Big( \Omega^\al D_\al  
\bar D^2 \, 
+ \, \bar \Omega_\beta \bar D^\beta  D^2 \, \Big) \,  
\left[  u \bar B^{\pm}
e^{\pm g V} B^{\pm} \right] \cdot \Gamma^{\rm 3d}_{\rm bdy}  \, , 
\end{gather}
and thus for the scale transformations by virtue of
(\ref{superconftr}) and (\ref{WD})
\begin{align}
\mu \frac{\pr}{\pr \mu} \Gamma^{\rm 3d}_{\rm bdy}
 = & -\frac{1}{2} \, \int\di^3x \di^2
\theta \di^2 \bar \theta \, u \, [ \bar B^{\pm}
e^{\pm g V} B^{\pm} ]  \cdot \Gamma^{\rm 3d}_{\rm bdy} \, . \label{scale22} 
\end{align}
The coefficient $u$ in (\ref{scale22}),
which is defined in (\ref{Zimmermann2}),
is related to $v$ in (\ref{Zimmermann2})
by $\N=4$, $d=3$ supersymmetry. Since $v$ vanishes, $u$ vanishes as
well if we assume $\N=4$, $d=3$ supersymmetry to be preserved upon 
quantization. In this case
we have immediately demonstrated conformal invariance, as
well as the absence of any anomalous dimensions.

Furthermore we may also show conformal invariance and the absence of
any anomalous dimensions for
the case of non-vanishing theta angle, in which we cannot set $u=0$  
based on $\N=4$ supersymmetry arguments.
For this purpose we 
derive the Callan-Symanzik equation corresponding to (\ref{scale22}). 
In analogy to 
(\ref{scale1}) leading to (\ref{CSE}), we write
\begin{gather}
( \mu \frac{\pr }{\pr \mu}  + \beta^g \pr_g
 - \gamma_B \N_B - \gamma_{Q_2} \N_{Q_2} - \gamma_V \N_V ) \,
\Gamma^{\rm 3d}_{\rm bdy} = 0 \, , \label{415}
\end{gather}
where
\begin{gather}
 {\cal N}_{Q_2} \equiv \int\!\di^3x \di^2 \theta \, Q_2
\frac{\delta}{\delta Q_2} \, + \, \int\!\di^3x \di^2 \bar \theta \,
\bar Q_2
\frac{\delta}{\delta \bar Q_2} \, , \quad {\cal N}_V \equiv \,
\int\!\di^3x \di^2 \theta \di^2 \bar \theta \, V \frac{\delta}{\delta
V} \,  
\end{gather}
and ${\cal N}_B$ as in (3.24).
Applying the derivative operators involving the beta and gamma functions 
in (\ref{415}) to $\Gamma^{\rm 3d}_{\rm bdy}$
generates operator insertions, for instance as in (\ref{cderiv}), and
further insertions as given by (\ref{Zimmermann2}).  
Comparing these insertions to the r.h.s.~of
(\ref{scale22}), we find that (\ref{415}) 
holds subject to the conditions
\begin{align} \label{conda2}
 \beta^g \pr_g z_B - 2 \gamma_B z_B - \frac{1}{2} (1-8\gamma_B) u = \, & 0 \, , \\
\beta^g - \gamma_V g = \, & 0 \, , \label{condb2}\\
\beta^g - 2 g \gamma_B - g \gamma_{Q_2} = \, & 0 \, ,
\label{condc2}\\
\beta^g\pr_g z_V - 2 z_V \gamma_V = \, & 0
\, , \label{condd2} \\
\beta^g\pr_g z_{Q_2} - 2 z_{Q_2}
\gamma_{Q_2}  = \, & 0 \,  \label{conde2}
\end{align}
on the insertion coefficients. 
Here (4.18) is a consequence of gauge invariance. 
(4.19), (4.20) and (4.21) are consequences of the fact that $v$, $w$  
and $y$ in (\ref{Zimmermann2}) vanish, respectively.
From (\ref{condb2}) and (\ref{condd2}) we obtain
\begin{gather} \label{v}
\gamma_V  (g\pr_g - 2) z_V = 0 \, .
\end{gather}
Since $z_V= \theta_{\rm YM}\, +$ ({higher order terms}),  $(g\pr g -2)z_V
\neq 0$ if $\theta_{\rm YM} \neq 0$.\footnote{We check explicitly in appendix
B.2 that the order $g^2$ contribution to $z_V$ vanishes.}  
Then (\ref{v}) implies that $\gamma_V$ vanishes to all orders in perturbation
theory. Furthermore for $\gamma_V=0$, (\ref{condb2}) implies $\beta^g=0$ to all
orders. From (\ref{conde2}) then follows $\gamma_{Q_2}=0$, from (\ref{condc2})
$\gamma_B=0$ and from (\ref{conda2}) $u=0$.\footnote{We confirm that
$\gamma_B=0$ at one-loop by an explicit calculation in appendix B.4.}
Thus all quantum corrections vanish and the theory is finite.

\section{Non-abelian theory} \label{Sec5}
\setcounter{equation}{0} 

\subsection{Construction of the Model}

We now construct the non-abelian generalization of the defect action
(\ref{abelianaction}), which preserves ${\cal N}=4$, $d=3$ supersymmetry,
using ${\cal N}=2$, $d=3$ superspace.  This model corresponds to a stack of D3
branes intersected by a D5-brane. The action was given in $\N=1$, $d=3$
language in \cite{Freedman}.

The bulk field content of the model in ${\cal N}=2$, $d=3$ superspace is as
follows.  There is a vector multiplet $V$ transforming as
\begin{align}
e^{V} \rightarrow e^{-i \Lambda^{\dagger}} e^{V} e^{i\Lambda} \, , 
\end{align}
where $\Lambda$ is a chiral multiplet. Both $V$ and $\Lambda$
are matrices in the fundamental representation of the $SU(N)$ Lie
algebra.   The  linear multiplet $\Sigma$ is defined by
\begin{align}
\Sigma \equiv \epsilon_{\alpha\beta} {\bar D}_{\alpha}(e^{-V}D_{\beta}e^V)
\,. \label{nonabelianlinear}
\end{align}
Under gauge transformations
\begin{align} 
\Sigma\rightarrow e^{-i\Lambda} \Sigma e^{i\Lambda}  \,.
\end{align}
While $\Sigma$ is not hermitian,  it 
satisfies 
\begin{align}
\Sigma^{\dagger} = e^{V} \Sigma e^{-V} \,.
\label{reality}
\end{align}
We also require three adjoint chiral superfields, $Q_1, Q_2$ and
$\Phi$, with the gauge transformation properties
\begin{align}
&Q_i \rightarrow  e^{-i\Lambda} Q_i e^{i\Lambda} \,, \\
&\Phi \rightarrow e^{-i\Lambda} \Phi e^{i\Lambda} -
e^{-i \Lambda}\partial_z e^{i \Lambda} \,.
\end{align}
Note that $\Phi$ is a connection in the $z$ direction and the
operator $\partial_z - \Phi$ transforms
covariantly.

For vanishing $\theta$ angle the ${\N=4}$, $d=4$ super Yang-Mills action can
then be written in $\N=2$, $d=3$ superspace as
\begin{align} \label{M3bulk}
S_{\rm bulk} =
\frac{1}{g^2} \int \! dz d^3x d^2 \theta d^2 {\bar\theta}\, {\rm Tr}
\left[
\Sigma^2 - \left(e^{-V}(\partial_z + \Phi^{\dagger})e^{V} + \Phi\right)^2
+ e^{-V}{\bar Q}_i e^V Q_i \right]  \nonumber \\
+\int\! dzd^3x d^2\theta\, {\rm Tr} \epsilon_{ij}Q_i[-\partial_z + \Phi, Q_j]
+ \int\!  dzd^3x d^2{\bar \theta} \, {\rm Tr} \epsilon_{ij}
{\bar Q}_i[\partial_z + {\bar\Phi}, {\bar Q}_j] \,.
\end{align}

The defect or boundary  component of the action which preserves
 ${\cal N}=4$, $d=3$ supersymmetry is, in ${\cal N}=2$, $d=3$ superspace,
\begin{align} \label{M3boundary}
S_{\rm boundary} = \int\! d^3x d^2 \theta d^2 {\bar \theta}\,  
({\bar B}_1 e^{V}B_1 + { B}_2 e^{-V}{\bar B}_2)  \nonumber\\
\hspace{3cm} 
+ \, \Big( \, \int\! d^3x d^2 \theta \, B_2 Q_2 B_1   + c.c. \, \Big)
\, . 
\end{align}
Here $B_1$ is in the fundamental and $B_2$ in the antifundamental
representation of the gauge group such that
\begin{align}
B_1 \rightarrow e^{-i\Lambda}B_1 \,, \qquad
B_2 \rightarrow B_2 e^{i\Lambda} \, , \qquad {\rm with} \; \Lambda \,
=\, \Lambda(\theta,\vec{x}, z=0) \,.
\end{align}
Together, $B_1$ and $B_2$ form a ${\cal N}=4$, $d=3$ hypermultiplet.

\subsection{Comments on Chern-Simons terms}

One may also include a $\theta_{ \rm YM}$ term in the bulk $\N=4$ theory.  
It is clear how to write such a term in  $\N=1$ or $\N=2$ 
superspace. These different superspace representations differ 
only by total derivatives which are
irrelevant in the absence of a boundary, such that $\N=4$
supersymmetry is preserved.
However in the presence of a boundary, the different superspace 
representations of the $\theta_{ \rm YM}$ term in the bulk
induce Chern-Simons
terms with different amounts of supersymmetry on the boundary, 
and some of the supersymmetry may be broken. 
For instance we may write a bulk $\theta_{\rm YM}$ term
in $\N=2$, $d=3$ superspace as
\begin{align} \label{Sthym}
S_{\theta_{\rm YM}} = \, \frac{\theta_{\rm YM}}{16 \pi^2} \, \int\! d^3x \int dz\,  {\rm Tr}
\Sigma \left(
e^{-V} (\partial_z+ \Phi^{\dagger}) e^{V} + \Phi \right)\, .
\end{align}
In the absence of a boundary, adding such a term does not break $\N=4$,
$d=4$ supersymmetry.
However in the presence of a boundary at $z=0$,  (\ref{Sthym}) 
induces a $\N=2$, $d=3$ 
Chern-Simons term, such that $\N=4$, $d=3$ supersymmetry is broken
down to $\N=2$, $d=3$.   In a non-abelian theory, this 
Chern-Simons term is of course not the whole story,  
since one must also add the contributions
of $\int {\rm tr} F\wedge F$ coming from infinity. 
We emphasize that the induced Chern-Simons term
need not have a quantized level (or quantized $\theta_{YM}$). 
In the non-abelian case,  
the $\N=2$, $d=3$ Chern-Simons
term does not have a local representation in $\N=2, d=3$ superspace.  
Instead one introduces an auxiliary direction $t$, 
such that the Chern-Simons term is given by \cite{Ivanov,Avdeev}
\begin{align} \label{SCS}
S_{\rm CS}  = \int\! d^3x \int_0^1 \! dt\, & {\rm Tr}\,  \Sigma(t) \, 
e^{-V(t)} \partial_t e^{V(t)} \, , \\ &
 {\rm with} \;\;\; \Sigma(t) \, \equiv \, 
{\bar D}^{\alpha}
\left(e^{-V(t)}D_{\alpha}e^{V(t)} \right) \, , \nonumber
\end{align}
where $V(t,\vec x)$ satisfies the boundary conditions $V(0,\vec x) =
0$ and $V(1,\vec x) = V(\vec x)$.
Note the resemblance between (\ref{SCS}) and (\ref{Sthym}).
The $\N=3$, $d=3$ completion of (\ref{SCS}) is obtained by adding 
the chiral terms
\begin{align}
S_{Q_2} \, = \, \int \!d^3x d^2\theta \, {\rm Tr} \, Q_2^2 + c.c. \, .
\end{align} 
However  there is no known $\N=4$, $d=3$ Chern-Simons term. 
Although one can have a bulk theta angle preserving 
$\N=4,d=4$ supersymmetry,  the introduction of a boundary when
$\theta_{YM} \ne 2\pi$ apparently allows at most $\N=3, d=3$ supersymmetry.
One might try to cancel the induced Chern-Simons term by adding
another boundary Chern-Simons term by hand. However, unlike the induced
Chern-Simons term,  the one added by hand has a quantized level and 
cannot restore $\N=4, d=3$ supersymmetry when $\theta_{YM} \ne 2\pi$.  

\subsection{Conformal Invariance}

By construction, the model given by (\ref{M3bulk}), (\ref{M3boundary}) 
coincides with
the model given in $\N=1$, $d=3$ language in \cite{Freedman}. We
discuss the relation between the $\N=2$ and the $\N=1$ formulation in
more detail in appendix C.
There are several lines of reasoning which suggest the quantum 
conformal invariance of the defect model. 
One argument is based on its 
role \cite{Freedman} as the holographic dual of a supergravity 
configuration discussed in \cite{Karch}.  
In this case one expects conformal invariance as a reflection of
the isometries of the supergravity configuration, in which 
a D5-brane  spans  an $AdS_4$ submanifold embedded in $AdS_5$.   
Another argument is that the model is a matrix model for a IIB 
five-brane compactified on a two-torus \cite{GanorSethi,sethicom}.
Quantum conformal invariance was proven in \cite{Freedman}
in the abelian case, in which there are no interactions in the bulk, but
remains to be proven in the non-abelian case.  We will not attempt a rigorous
proof here, but we give an argument for conformal invariance using our
$\N=2$, $d=3$ formulation of the model. The argument 
relies on the assumption that the classical ${\cal N}=4$, $d=3$ 
supersymmetry is unbroken by quantum corrections. 
Ideally, one would like to find a proof using a minimal number of
assumptions about which classical symmetries are preserved by quantum
corrections. 

Let us first consider the implications of the unbroken supersymmetry.  In
 ${\cal N} = 2$, $d=3$  superspace, one can make use of non-renormalization 
of the superpotential 
\begin{align}
W = B_2 Q_2 B_1 \, +\,  \int\! dz \,\epsilon_{ij}{\rm Tr}
Q_i[-\partial_z + \Phi, Q_j] \,.
\label{superpot}
\end{align}
The second term in the superpotential is rather surprising, since it 
gives the  bulk
kinetic terms for $Q$ which involve derivatives in the $z$ direction.    
Since the superpotential is not renormalized, these kinetic terms
are protected against quantum corrections.
In the absence of a defect, Lorentz
invariance would then imply the non-renormalization of all the kinetic terms
for $Q$.  In other words, the
K\"{a}hler potential term  ${\rm Tr}
e^{-V}{\bar Q}_i e^V Q_i$ in the bulk action is 
also not renormalized \footnote{ In fact this is another
way of proving non-renormalization
of the metric on the Higgs branch
in ${\cal N}=2$, $d=4$ gauge theories \cite{9603042}.}.
However, in the presence of a defect,  Lorentz
invariance is broken and this argument is not available to us.

Instead let us consider the ${\cal N}=4$, $d=3$ completion of the
superpotential.  The completion of the first term in (\ref{superpot})
is ${\bar B}_i e^V B_i$, which is therefore also not renormalized.
The ${\cal N}=4$, $d=3$ completion of the second term is
${\rm Tr} \, \left(e^{-V}(\partial_z + \Phi^{\dagger})e^{V} + \Phi\right)^2$.
Supersymmetry does not place any constraints on the renormalization
of the remaining term ${\rm Tr} \, (\Sigma^2 + e^{-V}{\bar Q}_i e^V Q_i)$.

Note that in the case of the ${\cal N} =4$ bulk theory without a
defect,  there are arguments for conformal invariance based on
the conservation of an R symmetry current in the same multiplet
as the stress-energy tensor \cite{West}.
Such arguments do not work when a defect is included.  In this case,
one can define a classically conserved R symmetry
current as
\begin{align} \label{JJ}
J^{M} = J_{(3)}^M + \int\! dz\, J_{(4)}^{M}
\end{align}
where $J_{(3)}$ is the contribution of defect terms, 
$J_{(4)}$ is the four dimensional R current, and
$M$ is a three dimensional Lorentz index with values $0,1,3$.
The charge associated with the current (\ref{JJ}) generates the R
symmetry transformations of the combined bulk and defect action. 
It is the lowest component of a supercurrent
${\cal J}^{\alpha\beta} = J^M \sigma_M^{\alpha\beta} + \dots$.  
Let us consider the possible anomalies of this current.
There is no reason to expect R symmetry to be broken by quantum
corrections.  However, unlike the four dimensional case, one can find 
a potential  anomaly multiplet such that the R current
is conserved, but the (three dimensional)  stress tensor has
non-zero trace.  For instance one could consistently write for the
supercurrent anomaly 
\begin{align}
{\bar D}^{\alpha}{\cal J}_{\alpha\beta} \sim \, \beta^g \, \int\! dz \, 
D_{\beta}{\bar D}{\bar D} \, {\rm Tr} (\Sigma^2 + e^{-V}{\bar Q}_i e^{V} Q_i) \,,
\label{anomaly}
\end{align}
The anomaly multiplet on the right hand side of (\ref{anomaly})
is gauge invariant.  It implies a non-zero trace for the stress energy
tensor,  however there is a conserved $R$ current since, with $
[ DD, \bar D \bar D] \,  = 16 \, \square_3 \, + \, 8i D \sigma^M
\bar D \pr_M$,
\begin{align}
\{ D^{\alpha}, {\bar D}^{\beta} \}
{\cal J}_{\alpha\beta} \, =  16 \, \beta^g \, 
\,  ( \square_3 \, + \, 8i D \sigma^M
\bar D \pr_M ) \, \int\! dz\, {\rm Tr} \, 
(\Sigma^2 + e^{-V}{\bar Q}_i e^{V} Q_i) \, . 
\end{align} 
The difference between the standard $\N=1$, $d=4$ anomaly $W^\al W_\al$ and
the $\N=2$, $d=3$ expression $ \int dz\, \bar D \bar D \Sigma^2$ is that the
latter is a gauge invariant term ($\Sigma^2$) 
chirally projected by $\bar D \bar D$. 
On the other hand, $W^\al W_\al$  cannot be written as the chiral
projection of a gauge invariant term.
It is well known that chirally projected anomalies may be
absorbed in a redefinition of the supercurrent ${\cal J}_{\al \beta}$, such
that R symmetry is manifestly conserved even if scale invariance is
broken \cite{PS,West}.

\begin{figure}[!ht]
\begin{center}
 \scalebox{.75}{\includegraphics{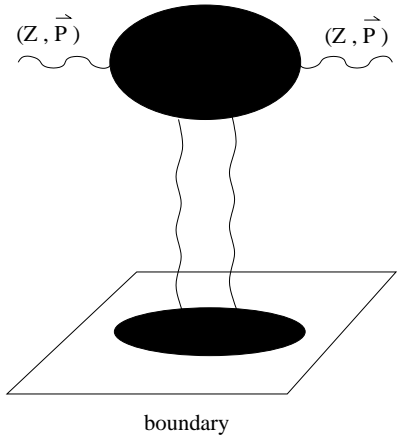}}
\caption{A $z$ dependent contribution to a bulk-bulk propagator.}
\label{zdep}
\end{center}
\end{figure}

The anomaly equation (\ref{anomaly}) is therefore permitted by ${\cal N}=4$,
$d=3$ supersymmetry. Nevertheless this anomaly must be absent for the following reason. 
Consider correlation functions of bulk
fields in the limit of large $z$, with  fixed momenta $\vec p$
parallel to the defect.  These receive the usual contributions
from diagrams involving only bulk fields.  Such contributions
are finite due to the finiteness of the ${\cal N}=4$, $d=4$ theory.  
Contributions from diagrams which involve bulk-defect
interactions (see  figure \ref{zdep})
are $z$ dependent and fall off with distance from the
defect.
Therefore any local counterterms from such diagrams would
have an explicit $z$ dependence. 
This means that the corresponding counterterms 
contributing to the action would be  of the schematic form
\begin{gather} \label{O}
\int\! d^3x d^2 \theta d^2 \bar \theta \, 
\int \! dz \, f(z)  \hat O(z, \vec{x}) \, ,
\end{gather}
with $f(z)$ falling off with the distance from the defect.
Clearly the anomaly in (\ref{anomaly}) is not consistent with
counterterms  of this form since its $z$ integrand 
has no explicit $z$ dependence as given by $f(z)$ in (\ref{O}). 
Therefore this anomaly must be absent and $\beta^g=0$ in (\ref{anomaly}).

Furthermore we may also rule out the possibility of any 
counterterms of the form (\ref{O}) and thus of any explicitly $z$ dependent 
anomalies contributing to (\ref{anomaly}): 
In addition to having the $z$ dependence described, 
any counterterm would have to be expressible as an integral of a 
local operator over all of ${\cal N} =2$, $d=3$ superspace.  
In other words, $z$ dependent counterterms as introduced in (\ref{O}) 
would have to take the asymptotic form
\begin{align}
\int\! d^3 x dz d^2 \theta d^2 {\bar \theta} \, 
z^{-s} \Lambda^{t}  {\hat O} \, , \label{cterm}
\end{align}
where $\Lambda$ is a mass scale, $s>0$ and $t \ge 0$.
The local operator $\hat O$ must therefore have dimension less than
$2$,  and be non-chiral.  It must also be gauge invariant
upon integration.  However there is no such operator available to 
us. The only apparent possibility is a bulk Chern-Simons term 
$\hat O= {\cal L}_{CS}$, but there is no such term 
preserving $\N=4$, $d=3$ supersymmetry.  
Furthermore there is no parity anomaly which could generate it.
We therefore conclude that the model is closed under the renormalization
group, requiring no additional $z$ dependent interactions.  
Since $\N=4,d=3$ supersymmetry also prevents 
renormalization of the defect interactions, we find that the
theory is conformal.

One can also consider a variant of the model given by (5.7), (5.8) 
in which a $\N=3$, $d=3$ Chern-Simons term localized at the defect
(not to be confused with the bulk Chern-Simons term discussed above)
is added by hand to the classical action.  This
still does not break conformal invariance at the quantum level.
The addition of a $\N=3$ Chern-Simons term breaks
$\N=4$ only in a very mild way. 
The $\N=3$ multiplets are the same as the $\N=4$ 
multiplets.  Furthermore in a renormalizable theory, $\N=3$ supersymmetry
automatically implies $\N=4$ supersymmetry for all terms except the 
Chern-Simons term.  There is now the apparent possibility of
generating a bulk Chern-Simons term of the type discussed in the
previous paragraph.  However such a term,  in addition to being
non-local,
has a chiral component and can therefore not arise from perturbative
quantum corrections. 
Of course such a $\N=3$ conformal model does not arise as a
holographic dual  
of the AdS configuration of \cite{Karch}, in which a D5-brane wraps an 
$AdS_4 \times S^2$ submanifold of $AdS_5 \times S^5$.  This configuration
has a $SU(2) \times SU(2)$ isometry corresponding to the $\N=4, d=3$
R symmetry group.  The $\N=3$ Chern-Simons term breaks this
symmetry to its diagonal $SU(2)$ subgroup. 

\section{Conclusions}\setcounter{equation}{0}

In this paper we have studied superconformal field theories coupled to
boundary or defect degrees of freedom in such a way as to preserve
the superconformal symmetries leaving the boundary or defect
locus invariant.  We constructed new abelian theories of this type 
which we showed to be conformally invariant. 
We have also given an argument that the non-abelian $\N=4$, $d=3$ model
constructed in \cite{Freedman} is conformally invariant by excluding
the possibility of counterterms which asymptotically fall off with the
distance from the boundary.
Our arguments rely greatly on the use of
$\N=2$, $d=3$ superspace, which was used to express both the
boundary/defect and bulk components of the action.  

Theories of the type discussed here are of interest in a variety
of systems. In the context of AdS/CFT duality, it
would be interesting to try to construct  a conformally invariant
boundary model in which there are four-dimensional conformal field
theories with different 
central charges on opposite sides of the
boundary. For this purpose the $\N=2$, $d=3$ superspace formalism and
the supersymmetric boundary conditions introduced in this paper may
prove to be useful.
The AdS dual of such a theory would presumably consist
of two AdS spaces of different curvature separated by an AdS
sub-manifold \cite{Karch}.  
It may also be interesting to consider defect theories that arise from
orbifolds of the $S^5$ in the AdS configuration of \cite{Karch}.
Orbifolding the $S^5$ in the conventional $AdS_5 \times S^5$
background gives string theory duals of large $N$ conformal field 
theories with less than $\N=4$ supersymmetry \cite{9802183}.
In the background of \cite{Karch} which leads to the defect
model of \cite{Freedman},  a D5-brane wraps an $AdS_4 \times S^2$
submanifold of $AdS_5 \times S^5$.  By orbifolding the $S^5$ one
could concievably obtain large $N$ conformal defect models with less
or no supersymmetry.   

Moreover we expect further interesting non-supersymmetric
boundary or defect 
conformal theories of this type to exist. A simple example is for
instance  the minimal non-supersymmetric model given by
\begin{align} 
S = \int\! dzd^3x\, F^{\mu \nu} F_{\mu \nu} \, + \, 
\int_{z=0}\! d^3x \,{\bar \Psi} 
\gamma^M( \partial_M - i A_M) \Psi \, ,
\end{align} 
with a boundary fermion $\Psi$ coupled to a bulk gauge field $A_\mu$. 
In the abelian case this model
is conformal, as may be seen straightforwardly by using the
method of sections 3 and 4 without supersymmetry: Instead of the
supercurrent anomaly, a conformal transformation of the boundary or
defect  1PI action gives now an insertion of the trace of the stress-energy
tensor. In this case conformal invariance is potentially broken by
insertions of the form
\begin{gather}
[ T^M{}_M ] \cdot \Gamma^{\rm 3d} \, = \, [ \, u\,  {\bar \Psi} 
\gamma^M( \partial_M - i A_M) \Psi \, + \, v\, \varepsilon^{MNP}A_M
F_{NP} ] \cdot \Gamma^{\rm 3d} \, .
\end{gather}
Again $v$ has to vanish since $\varepsilon^{MNP}A_M
F_{NP}$ is not gauge invariant. $u$ gives rise to an anomalous
dimension for the fermion field, such that we have again a conformal
field theory with an anomalous dimension for the boundary field.
Such theories may arise also in the context of
critical phenomena of systems with interacting boundaries or
defects.

\vspace{5cm}

\bigskip

{\bf \large Acknowledgements}

\medskip

We are grateful to Luis Alvarez-Gaum\'{e}, Jos\'{e} Barb\'{o}n, David Berman,
Oliver DeWolfe, Sergio Ferrara, Dan Freedman and Savdeep Sethi for useful 
comments on the first version of this paper.

\smallskip

Our research is funded by the DFG (Deutsche Forschungsgemeinschaft)
within the Emmy Noether programme, grant ER301/1-2.

\newpage
\appendix
\section{Linear multiplet and decomposition of $\Psi$}

\subsection{Component expansion of the 3d $\N=2$ linear multiplet} 
\label{App1a}

For the expansion of the (abelian) linear multiplet $\Sigma
\equiv\ts \frac{1}{2i} \ve^{{\a}\b} \bar D_{ \a} D_\b V$, we have to
differentiate twice the 3d vector superfield $V$ which in Wess-Zumino gauge is
given by
\begin{align}
V =&\, - \theta \s^2 \bar\theta \rho - \theta \s^M \bar\theta v_M
+ i (\theta\theta) \bar\theta \bar\l - i (\bar\theta\bar\theta) \theta \l
+ \frac{1}{2} (\theta\theta)(\bar\theta\bar\theta) D\,.
\end{align}
The real scalar $\rho$ stems from the second component of the four
vector $v_\m$, i.e.\ \mbox{$\rho \equiv v_2$}. In chiral 
coordinates 
\begin{align}
y^M = x^M+i \theta \s^M \bar\theta \,, \quad M=0,1,3 \,,
\end{align}
the derivative $\bar D_{\a}$ takes the simple form $\bar D_{\a}= - \bar
\partial_{\a}$. For $D_\b V(y,\theta,\bar\theta)$ we find
\begin{align}
  D_\b V(y,\theta,\bar\theta)=&-\s^2_{\b\g} \bar \theta^\g \rho(y) -\s^M_{\b\g}
  \bar \theta^\g v_M (y) + 2i \theta_\b \bar \theta^\a \bar \l_\a(y)-
  i(\bar\theta\bar\theta) \l_\b(y)  \nonumber\\
  &+ \theta_\b (\bar\theta\bar\theta) (D(y) + i \partial^M v_M (y) ) -
  i(\bar\theta\bar\theta) (\s^M \bar \s^N \theta)_\b \partial_M v_N(y)\\
  &- i(\bar\theta\bar\theta) (\s^M \bar \s^2 \theta)_\b \partial_M \rho(y) +
  (\theta\theta)(\bar\theta\bar\theta) \s^M_{\b\g} \partial_M \bar \l^\g\,.
  \nonumber
\end{align}
Using the identity $\s^M \bar \s^N = \eta^{MN} -i \s^{MN}, 
\eta^{MN}={\rm diag}(1,-1,-1)\,,$ we end up with
\begin{align}
  \Sigma(y,\theta,\bar\theta)&=\rho + \theta \bar\l+\bar\theta\l +i \theta
  \bar\theta D + \frac{i}{2} \bar\theta \s^{MN}  \theta F_{MN} + i
  (\theta\theta) \bar\theta \s^M \partial_M \bar \l \,,
\end{align}
where the field strength $F_{MN}$ is given by
\begin{align}
F_{MN}\equiv\partial_M v_N -\partial_N v_M \,.
\end{align}
Further expansion leads to
\begin{align} \label{linear3D}
  \Sigma(x,\theta,\bar\theta)=&\,\rho+ \theta \bar\l+\bar\theta\l +i
  \theta \bar\theta D + \frac{1}{2} \bar\theta \s_{K} \theta
  \varepsilon^{MNK} F_{MN} \nonumber\\
  &+ \frac{i}{2} (\theta\theta) \bar\theta \bar \s^M 
  \partial_M \bar \l + \frac{i}{2} (\bar\theta\bar\theta) \theta \s^M 
  \partial_M \l - \frac{1}{4} (\theta\theta)(\bar\theta\bar\theta)
  \square_3 \rho \,.
\end{align}

\subsection{Detailed calculation of $\Psi$ at $\thetasl=0$}  \label{App1b}

In this appendix we show some details of the calculation of $\Psi=\Phi' + i
\sqrt{2} \theta^\a_2 W'_\a + \theta_2 \theta_2 G'$ at $\thetasl=0$.  Since the
chiral superfield $\Phi'$ is in the adjoint representation, in the abelian
case the auxiliary field $G'$ is given by
\begin{align}
  G'(\tilde y,\theta_1) &\equiv \int\! d^2 \bar \theta^1 \, \bar\Phi' (\tilde
  y - 2 i \theta_1 \s \bar \theta^1, \theta_1) \nonumber\\
  &= F'{}^*(\tilde y) - {i}{\sqrt{2}} \theta_1 \s^m \partial_m \bar\psi'(\tilde
  y) -(\theta_1\theta_1) \square \phi'{}^*(\tilde y) \,.
\end{align} 
Substituting the coordinate transformation (\ref{inversetrafo2}) and the
component expansions of $\Phi'$, $W'_\a$, and $G'$ into
(\ref{N=1expansion}), we find for
$\Psi\vert_{\thetasl=0}(\tilde y^\m, \theta, \bar \theta)$ \footnote{
$SL(2,\mathbb{R})$ invariant products are defined in the following way:
$\theta^2 \equiv \theta^\a \theta_\a$, 
$\bar \theta^2 \equiv \bar\theta_\a\bar\theta^\a $,
$\theta \bar \theta \equiv
\theta^\a \bar \theta_\a = \bar \theta^\a\theta_\a$.}
\begin{align}
  \Psi\vert_{\thetasl=0} =&\,\phi' + \frac{1}{\sqrt{2}}\theta(\psi'+i\l') +
  \frac{1}{\sqrt{2}} \bar\theta(\psi'-i\l') + \frac{1}{4} \theta\theta
  (F'-F'{}^*-\sqrt{2} D') \nonumber\\ &- \frac{1}{2} \theta\bar\theta (F'+F'{}^*)
  + \frac{1}{4} \bar\theta\bar\theta (F'{}^*-F'-\sqrt{2} D') + \frac{1}{4}
  \sqrt{2} \bar \theta \s^{\m\n} \theta F'_{\m\n}\\& + \frac{i}{2}
  \frac{1}{\sqrt{2}} (\bar \theta\bar \theta) \theta \s^\m \partial_\m (\bar
  \psi' - i\bar \l') +\frac{i}{2} \frac{1}{\sqrt{2}} (\theta \theta)
  \bar\theta \s^\m \partial_\m (\bar\psi' +i \bar \l') - \frac{1}{4} (\theta
  \theta)( \bar \theta\bar \theta) \square \phi'{}^* \nonumber \,.
\end{align}

We now use the redefinitions (\ref{redef1}) - (\ref{redef2}) and expand
$\Psi\vert_{\thetasl=0}$ in $\tilde y^\m = x^\m + i \theta \s^2 \bar \theta
\d^\m_2 $. We obtain the component expansion of Eqn.\ (\ref{decomposition2})
\begin{align}
  \Psi\vert_{\thetasl=0}(x,\theta,\bar\theta) =&\, {\rm Re}\,\phi + i
  \frac{1}{\sqrt{2}} \rho + \frac{1}{\sqrt{2}} \theta(\psi+i\bar\l) +
  \frac{1}{\sqrt{2}} \bar\theta(\bar\psi+i\l)
  -  \frac{1}{\sqrt{2}} \theta\bar\theta D\nonumber\\
  &+ \frac{1}{2} \theta\theta F + \frac{1}{2} \bar\theta\bar\theta F^* - \theta \s^{M} \bar \theta
  \partial_M {\rm Im}\,\phi + \frac{1}{\sqrt{2}} \theta \s^M\bar\theta
  \partial_2 v_M  \nonumber\\
  &+ \frac{1}{2} \frac{i}{\sqrt{2}} \bar \theta \s_{K} \theta
  \varepsilon^{MNK} F_{MN} + \frac{i}{2} \frac{1}{\sqrt{2}} (\bar \theta\bar
  \theta) \theta \s^M \partial_M ( \bar\psi -i \l)\\
  &+\frac{i}{2} \frac{1}{\sqrt{2}} (\theta \theta) \bar\theta \bar\s^M
  \partial_M (\psi -i \bar \l) - \frac{1}{4}
  (\theta \theta) (\bar\theta\bar\theta) \square_3 \phi'{}^* \,,\nonumber\\
  & - \frac{1}{\sqrt{2}} \theta \s^2 \bar\theta \partial_2 \rho +
  \frac{i}{\sqrt{2}} (\theta \theta) \bar\theta \partial_2 \bar \l -
  \frac{i}{\sqrt{2}} (\bar\theta \bar\theta)\theta \partial_2 \l + \frac{1}{2}
  \frac{1}{\sqrt{2}} ( \theta\theta)(\bar\theta\bar\theta) \partial_2 D
  \nonumber\\
  =&\,\frac{1}{2} (\Phi + \bar \Phi +\sqrt{2}
  \partial_2V) + i \frac{1}{\sqrt{2}}\Sigma \,,\nonumber
\end{align}
with $\square_3\equiv \partial^M \partial_M$. The expansion yields terms
involving the transverse derivative $\partial_2$. While some of these terms
cancel higher order terms, the term $\partial_2 {\rm Re}\,\phi'$ is absorbed in
the definition of $D$, the auxiliary field of the linear field $\Sigma$. The
remaining ones form the superfield $\partial_2 V$.

\section{ \label{1loopapp}  Feynman rules and one-loop contributions}
\setcounter{equation}{0}

\subsection{Free 3d and 4d propagators}

We use a $\N=2$, $d=3$ superspace formulation for calculating quantum 
corrections. The use of superspace Feynman rules automatically guarantees the
cancellation of delta function
singularities and interactions at the boundary, 
which are explicitly present in the component formulation \cite{Mirabelli}. 

When both ends are pinned on the boundary, the free 4d chiral  and 
gauge vector propagators are given by
\begin{align} \label{propQV}
  G^{\rm bdy\rightarrow bdy}_{Q} (\vec p)=\frac{\delta^7_{zz'}}{2 \vert
    \vec p \vert}, \quad G^{\rm bdy\rightarrow bdy}_{V}(\vec p) =
  \frac{-\delta^7_{zz'}}{2 \vert \vec p \vert} \,.
\end{align}
where
\begin{align}
\delta^7_{zz'} = \delta^3 (\vec x - \vec x') \delta^2(\theta-\theta')
\delta^2(\bar \theta- \bar \theta')\,.
\end{align}
These expressions are the $\N=2$, $d=3$ version of results obtained in
\cite{Georgi} for the 4d/5d case in component form. The power of the
momentum in the denominator is reduced as compared to standard
three-dimensional propagators, which makes Feynman graphs potentially
more divergent than in a pure three-dimensional theory.
The standard free 3d propagators for the chiral boundary or defect
fields are
\begin{align} \label{propB}
G_{B_i}(\vec p)=\frac{\delta^7_{zz'}}{\vec p^2}, \quad i=1,2 \,.
\end{align}

\subsection{One-loop contribution to the Chern-Simons term}

We calculate the one-loop contribution to the coefficient $z_V$ of the
Chern-Simons term within the BPHZ approach. The calculation is analogous to
the standard calculation of the one-loop contribution to the 4d gauge
propagator given for instance in \cite{West,Grisaru}. The BPHZ approach allows
to show in a simple way that the one-loop contribution to the Chern-Simons
term in $d=3$ vanishes due to supersymmetry.

\begin{figure} 
\begin{center}
\scalebox{.75}{\includegraphics[-20,580][566,793]{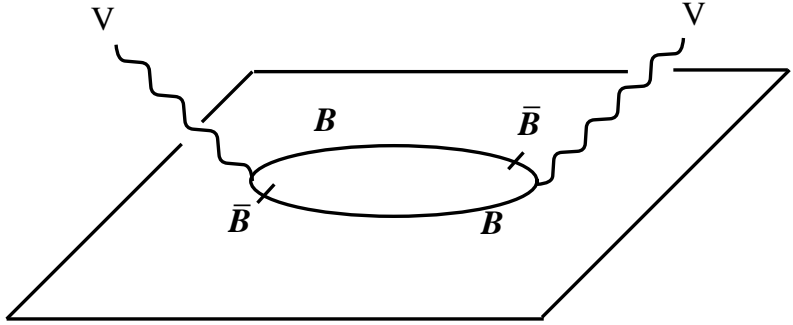}}
\caption{One loop contribution to the bulk propagator.}\label{diagram}
\end{center}
\end{figure}
The action principle of the BPHZ approach gives
\begin{gather}
\frac{\delta^2}{\delta V(1) \delta V(2) }\Gamma^{\rm 1PI} \Big|_{\rm 1-loop}
= \frac{\delta^2}{\delta V(1) \delta V(2)} 
\Gamma^{\rm eff}_{\rm 1-loop} \, ,
\end{gather} 
where the contributions relevant here are
\begin{align}
\Gamma^{\rm 1PI}_{\rm 1-loop} = 
& \, g^2 \int \frac{d^3 p}{(2\pi)^3} 
 \frac{d^3 k}{(2\pi)^3}\,
d^4 \theta \, d^4 \theta' \, V(- \vec{p}, \theta) \frac{\bar D^2 D^2}{16}
\frac{\delta^7_{zz'}}{(\vec p + \vec k)^2} \frac{D^2 \bar D^2}{16}
\frac{\delta^7_{zz'}}{\vec k^2} \, V(\vec{p}, \bar \theta)\, 
+ ... \, , \label{ce}\\
\Gamma^{\rm eff}_{\rm 1-loop} = & \, z_V^{(1)} \int \! d^3x d^4\theta \, V \Sigma
\, + ...\, .  
\end{align}
(\ref{ce}) corresponds to the graph shown in Fig.~2 with the external
legs removed as appropriate for a 1PI contribution. This ensures that
this 1PI contribution is three-dimensional.
Performing the two functional derivatives with respect to $V$ we
obtain
\begin{align}
\Gamma^{(2)}_{\rm 1-loop} & = z_V{}^{(1)} \varepsilon^{\alpha \beta} D_\al
\bar D_\beta ( \theta^2 \bar \theta^2) \, , \quad  \label{kkk}\\
{\rm where} \qquad & \Gamma^{(2)}_{\rm 1-loop} \, \equiv \, \frac{\delta^2}{\delta V(1) \delta
V(2) } \Gamma^{\rm 1PI} \Big|_{\rm 1-loop} = \, g^2 \, \int \! \frac{d^3 k}{(2\pi)^3} 
\, \frac{1}{(\vec p + \vec k)^2 \vec k^2}  \, .
\end{align}
Applying $\varepsilon_{\gamma \delta}D^\gamma \bar D^\delta$ 
to both sides of (\ref{kkk})
then gives
\begin{gather}
z_V^{(1)} \, = \, 0 \, ,
\end{gather}
since $\Gamma^{(2)}_{\rm 1-loop}$ is independent of the Grassmann variables.

\subsection{$\N=2$ model - One loop correction to the boun\-dary propagator 
  for the defect field $B$} \label{A.3} 

\begin{figure}[!h]
\begin{center}
\scalebox{.75}{\includegraphics[-95,620][566,793]{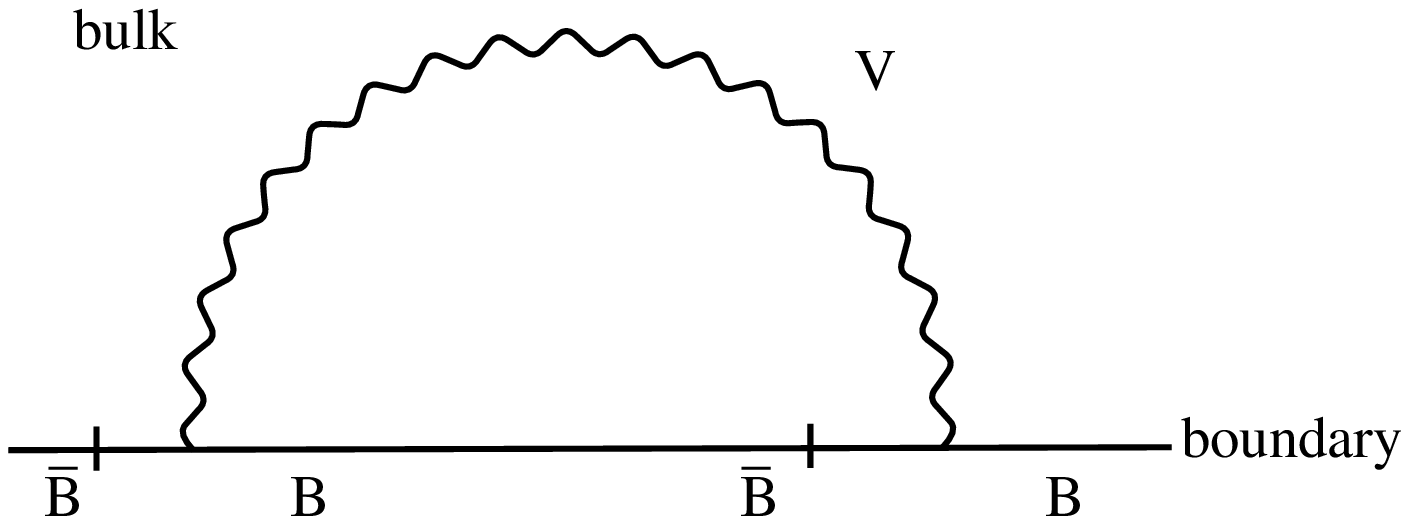}}
\caption{One loop contribution to the boundary propagator.}
\end{center}
\end{figure}

A standard one-loop calculation gives a contribution to the 1PI action.
By virtue of \mbox{Eqns.\ }(\ref{propQV}) and (\ref{propB}), we find
\begin{align} \label{loopcontr1}
  \Gamma^{\rm 1PI}_{B\bar B} = g^2 \int \frac{d^3 p}{(2\pi)^3}\frac{d^3
    k}{(2\pi)^3}\, d^4 \theta \, B(\vec{p}, \theta, \bar \theta) \frac{-1}{2
    \vert \vec p + \vec k \vert} \frac{1}{\vec k^2 } \bar B (\vec{p},
  \theta,\bar \theta) \,
\end{align}
for the super-Feynman graph of Fig.~3. 
Since $\Gamma^{\rm 1PI}_{B\bar B}$ is logarithmically divergent,
it follows that $\gamma_B \neq 0$ in general.

\subsection{$\N=4$ model - Additional one loop correction to 
  the boun\-dary propagator for the defect field $B$} 

In addition to the expression (\ref{loopcontr1}), in the $\N=4$ model we find
the following one-loop contribution to the 1PI action 
$\Gamma^{\rm 1PI}_{B\bar B}$, see Fig.~\ref{Fig4}:
\begin{figure}[!ht]
\begin{center}
 \scalebox{.75}{\includegraphics[-95,620][566,793]{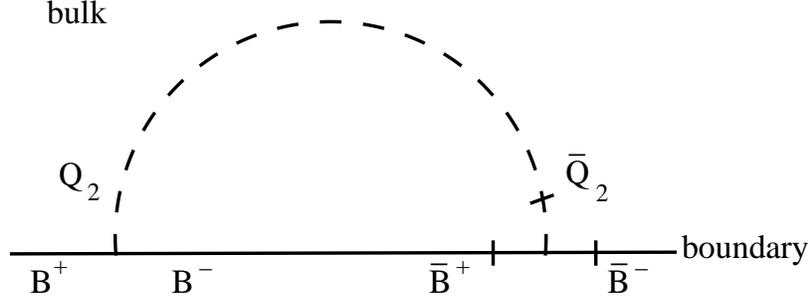}}
\caption{One loop contribution to the boundary propagator.}
\label{Fig4}
\end{center}
\end{figure}
\begin{align}
  \Gamma^{\rm 1PI}_{B\bar B} =&\,g^2 \int \frac{d^3 p}{(2\pi)^3}\frac{d^3 k}{(2\pi)^3}\, d^4 
   \theta \, d^4 \theta' \,\nonumber\\
  &\times B(- \vec{p}, \theta,\bar \theta) \left( - \frac{\bar
      D^2}{4} \right) \left( -\frac{D'{}^2}{4} \right) \frac{\delta^7_{zz'}}{2
    \vert \vec p + \vec k \vert}
  \frac{\delta^7_{zz'}}{\vec k^2 } \bar B(\vec{p}, \theta, \bar \theta) \nonumber\\
  =&\,g^2 \int \frac{d^3 p}{(2\pi)^3}\frac{d^3 k}{(2\pi)^3}\, d^4 \theta
  \, B(\vec{p}, \theta,\bar \theta) \frac{1}{2 \vert \vec p + \vec k \vert}
  \frac{1}{\vec k^2 } \bar B(\vec{p}, \theta, \bar \theta) \,.
\end{align}
We see that this exactly cancels the contribution (\ref{loopcontr1})
such that $\g_B=0$ at least to one loop.

\section{ The $\N=4$ model of (\ref{M3boundary}) in $\N=1$, $d=3$
language} 

\setcounter{equation}{0}

By construction, the model given by (\ref{M3bulk}), (\ref{M3boundary})
is the $\N=2$, $d=3$ formulation of the 
model constructed in \cite{Freedman} in $\N=1$, $d=3$ notation.
To demonstrate this we note that the gauge field $V$ decomposes into 
\cite{Ivanov}
\begin{gather}
V^a T^a  \, = \, - i \, \Gamma^{\al a} \tilde \theta_{2 \alpha} T^a 
\, + \, {\ts \frac{1}{2}} \, i \, \tilde \theta_2{}^\al \tilde \theta_{2
\al} b^a T^a \, 
\end{gather}
under $\N=1$, $d=3$ supersymmetry. Here $\tilde\theta_2$ is defined in
(2.2). The kinetic term in
(\ref{M3boundary})  decomposes into
 \begin{align} \label{f}  \, 
\int \! d^3x d^2 \tilde \theta_1 \big( (\overline{\nabla q_i}) \nabla q^i \, +
b q^i \bar q_i \big) \,,
\quad i=1,2 \quad,
\end{align}
where the $\N=1$, $d=3$
 superfield $q_i$ is defined as the lowest component in a
$\tilde\theta_2$ expansion of $B_i$, $q_i=B_i
 \vert_{\tilde\theta_2=0}$. 
The covariant derivative in (\ref{f}) 
is given by $\nabla \, =\, D- i \Gamma^a T^a$ with $D$ the $\N=1$,
 $d=3$ derivative. The kinetic term in
(\ref{f}) coincides with the boundary kinetic term given in 
(4.27) of \cite{Freedman} and $b q^i \bar q_i$ contributes to the
 superpotential term (4.29) of \cite{Freedman}.  $b$ contains the
 auxiliary field $D$ defined in (2.21) above and thus $\pr_2 {\rm Re}
 \phi'$,  which coincides with one of the three
 hypermultiplet scalar normal derivatives 
 $D_6 X_H{}^A$ which appear in the superpotential (4.29) of 
 \cite{Freedman}. 
The superpotential term in (\ref{M3boundary}) contains a complex
 auxiliary field $f$ and thus the two remaining hypermultiplet scalar 
derivatives of the form $D_6 X_H{}^A$.


\begin{thebibliography}{99}

\bibitem{Cardy} J.L. Cardy, Nucl.\ Phys.\ B {\bf 240} [FS12](1984) 514.

\bibitem{OsbornMcAvity} D.M.~McAvity and H.~Osborn,
Nucl.\ Phys.\ B {\bf 455} (1995) 522, [arXiv:cond-mat/9505127];\\
D.M.~McAvity and H.~Osborn,
Nucl.\ Phys.\ B {\bf 406} (1993) 655.  

\bibitem{Sethi}
S.~Sethi,
Nucl.\ Phys.\ B {\bf 523}, 158 (1998)
[arXiv:hep-th/9710005].

\bibitem{GanorSethi}
O.~J.~Ganor and S.~Sethi,
JHEP {\bf 9801} (1998) 007
[arXiv:hep-th/9712071].

\bibitem{KapustinSethi}
A.~Kapustin and S.~Sethi,
Adv.\ Theor.\ Math.\ Phys.\  {\bf 2}, 571 (1998)
[arXiv:hep-th/9804027].



\bibitem{Karch} A.~Karch and L.~Randall, 
A.~Karch and L.~Randall,
JHEP {\bf 0106} (2001) 063
[arXiv:hep-th/0105132].

\bibitem{Freedman} O. DeWolfe, D.Z. Freedman and H. Ooguri, 
[arXiv:hep-th/0111135].

\bibitem{Hori} K.~Hori, 
[arXiv:hep-th/0012179].

\bibitem{sethicom} S.~Sethi, private communication.

\bibitem{HG} N. Arkani-Hamed, Th. Gregoire, J. Wacker, 
[arXiv:hep-th/0101233].

\bibitem{Hebecker} A.~Hebecker,
[arXiv:hep-ph/0112230].


\bibitem{HananyWitten} A.~Hanany and E.~Witten,
Nucl.\ Phys.\ B {\bf 492} (1997) 152.
[arXiv:hep-th/9611230].


\bibitem{Lechtenfeld}
E.~Ivanov, S.~Krivonos and O.~Lechtenfeld,
Phys.\ Lett.\ B {\bf 487} (2000) 192
[arXiv:hep-th/0006017].


\bibitem{Lykken} J.D. Lykken, 
[arXiv:hep-th/9612114].

\bibitem{Nishino} H. Nishino and S.J. Gates, 
Int.\ J.\ Mod.\ Phys.\ A {\bf 8} (1993) 3371.


\bibitem{Aharony} 
O.~Aharony, A.~Hanany, K.~A.~Intriligator, N.~Seiberg and M.~J.~Strassler,
Nucl.\ Phys.\ B {\bf 499} (1997) 67
[arXiv:hep-th/9703110].


\bibitem{PS}
T.~E.~Clark, O.~Piguet and K.~Sibold,
Nucl.\ Phys.\ B {\bf 143} (1978) 445; \\
T.~E.~Clark, O.~Piguet and K.~Sibold,
Nucl.\ Phys.\ B {\bf 172} (1980) 201; \\
T.~E.~Clark, O.~Piguet and K.~Sibold,
Nucl.\ Phys.\ B {\bf 169} (1980) 77.

\bibitem{ERS}
J.~Erdmenger, C.~Rupp and K.~Sibold,
Nucl.\ Phys.\ B {\bf 530} (1998) 501 
[arXiv:hep-th/9804053]; \\
J.~Erdmenger and C.~Rupp,
Annals Phys.\  {\bf 276} (1999) 152
[arXiv:hep-th/9811209]; \\
J.~Erdmenger, C.~Rupp and K.~Sibold,
Nucl.\ Phys.\ B {\bf 565} (2000) 363
[arXiv:hep-th/9907169].

\bibitem{BPHZ}
O.~Piguet and S.~Sorella, {\it Algebraic Renormalization}. 
Springer Verlag, Berlin 1995; \\ J.~Collins, {\it Renormalization}. 
Cambridge University Press 1984.

\bibitem{Zimmermann}
W.~Zimmermann,
Annals Phys.\  {\bf 77} (1973) 536
[Lect.\ Notes Phys.\  {\bf 558} (1973) 244]; \\
T.~E.~Clark and J.~H.~Lowenstein,
Nucl.\ Phys.\ B {\bf 113} (1976) 109.


\bibitem{Piguet}
N.~Maggiore, O.~Piguet and M.~Ribordy,
Helv.\ Phys.\ Acta {\bf 68} (1995) 264
[arXiv:hep-th/9504065].

O.~M.~Del Cima, D.~H.~Franco, J.~A.~Helayel-Neto and O.~Piguet,
JHEP {\bf 9802} (1998) 002
[arXiv:hep-th/9711191].


\bibitem{noCS2}
R.~Brooks and S.~J.~Gates,
Nucl.\ Phys.\ B {\bf 432} (1994) 205
[arXiv:hep-th/9407147].

\bibitem{Kapustin} A.~Kapustin and M.J.~Strassler, 
JHEP {\bf 9904} (1999) 021
[arXiv:hep-th/9902033].


\bibitem{ZK}
B.~M.~Zupnik and D.~V.~Khetselius,
Sov.\ J.\ Nucl.\ Phys.\  {\bf 47} (1988) 730
[Yad.\ Fiz.\  {\bf 47} (1988) 1147].

\bibitem{KLL}
H.~C.~Kao, K.~M.~Lee and T.~Lee,
Phys.\ Lett.\ B {\bf 373} (1996) 94
[arXiv:hep-th/9506170].

\bibitem{Ivanov} E.A.~Ivanov, 
Phys.\ Lett.\ B {\bf 268} (1991) 203.


\bibitem{Avdeev} L.V.~Avdeev, D.I.~Kazakov and I.N.~Kondrashuk,
Nucl.\ Phys.\ B {\bf 391} (1993) 333, [arXiv:hep-th/9302068].



\bibitem{9603042}
P.~C.~Argyres, M.~Ronen Plesser and N.~Seiberg,
Nucl.\ Phys.\ B {\bf 471} (1996) 159
[arXiv:hep-th/9603042].


\bibitem{West} P.~West, ``Introduction to supersymmetry and supergravity'',
Second Edition, World Scientific Publishing, Singapore 1990.



\bibitem{9802183}
S.~Kachru and E.~Silverstein,
Phys.\ Rev.\ Lett.\  {\bf 80} (1998) 4855
[arXiv:hep-th/9802183].

\bibitem{Mirabelli}
E.~Mirabelli and M.~Peskin, 
Phys.~Rev.~D {\bf  58} (1998) 065002, [arXiv:hep-th/9712214]. 



\bibitem{Georgi} H.~Georgi, A.K.~Grant and G.~Hailu, 
Phys.\ Lett.\ B {\bf 506} (2001) 207, [arXiv:hep-ph/0012379].


\bibitem{Grisaru} S.J.~Gates, M.T.~Grisaru, M.~Ro$\check{\rm c}$ek, and 
W.~Siegel,
Front.\ Phys.\ {\bf 58} (1983) 1,
[arXiv:hep-th/0108200].  


\end{thebibliography}
\end{document}